\documentclass[a4paper,fleqn,usenatbib]{mnras}
\usepackage{amsmath}
\usepackage{bm}
\usepackage{txfonts}
\usepackage{multicol}

\usepackage[T1]{fontenc}
\usepackage{ae,aecompl}

\usepackage{graphicx}	
\usepackage{grffile}

\usepackage{epsf}

\AtBeginDocument{}%
\begin{document}
	\newcommand{\msun}{\,{\rm M}_{\odot}}
	\newcommand{\rsun}{R_{\odot}}
	\newcommand{\kms}{\, {\rm km\, s}^{-1}}
	\newcommand{\cm}{\, {\rm cm}}
	\newcommand{\gm}{\, {\rm g}}
	\newcommand{\erg}{\, {\rm erg}}
	\newcommand{\kel}{\, {\rm K}}
	\newcommand{\pc}{\, {\rm pc}}
	\newcommand{\kpc}{\, {\rm kpc}}
	\newcommand{\mpc}{\, {\rm Mpc}}
	\newcommand{\seg}{\, {\rm s}}
	\newcommand{\kev}{\, {\rm keV}}
	\newcommand{\hz}{\, {\rm Hz}}
	\newcommand{\etal}{et al.\ }
	\newcommand{\yr}{\, {\rm yr}}
	\newcommand{\gyr}{\, {\rm Gyr}}
	\newcommand{\eq}{eq.\ }
	\newcommand{\amunit}{\msun {\rm AU^2/yr}}
	\def\arcsec{''\hskip-3pt .}
	
	\def\gapprox{\;\rlap{\lower 3.0pt                       
			\hbox{$\sim$}}\raise 2.5pt\hbox{$>$}\;}
	\def\lapprox{\;\rlap{\lower 3.1pt                       
			\hbox{$\sim$}}\raise 2.7pt\hbox{$<$}\;}
	
	\newcommand{\figsizeFour}{9.0cm}
	
	\newcommand{\figwidthSingle}{14.0cm}
	\newcommand{\figwidthDouble}{7.50cm}
	
	\newcommand{\figbig}{\figwidthSingle}
	\newcommand{\figsmall}{\figwidthDouble}

	\newcommand{\figappend}{\figwidthSingle}
	
	\newcommand{\reqOne}[1]{equation~(\ref{#1})}
	\newcommand{\reqTwo}[2]{equations~(\ref{#1}) and~(\ref{#2})}
	\newcommand{\reqNP}[1]{equation~\ref{#1}}
	\newcommand{\reqTwoNP}[2]{equations~\ref{#1} and~\ref{#2}}
	\newcommand{\reqTo}[2]{equation~(\ref{#1})-(\ref{#2})}
	\newcommand{\rn}[1]{(\ref{#1})}
	\newcommand{\ern}[1]{equation~(\ref{#1})}
	\newcommand{\be}{\begin{equation}}
	\newcommand{\ee}{\end{equation}}
	\newcommand{\ff}[2]{{\textstyle \frac{#1}{#2}}}
	\newcommand{\ben}{\begin{enumerate}}
		\newcommand{\een}{\end{enumerate}}
	\newcommand{\tred[1]}{\textcolor{red}{#1}}
	\newcommand{\tgreen[1]}{\textcolor{green}{#1}}
\newcommand{\BK}[1]{{\color{red}\bf BK: #1}}

\title[Stellar discs in galactic nuclei]{A numerical study of stellar discs in galactic nuclei}

\author[Panamarev \& Kocsis]
  {Taras Panamarev$^{1,2,3}$\thanks{Corresponding author email: panamarevt@gmail.com}, Bence Kocsis$^1$ \\
   \\
      $^1$ Rudolf Peierls Centre for Theoretical Physics, Parks Road, OX1 3PU, Oxford, UK\\
      $^2$ Fesenkov Astrophysical Institute, Observatory 23, 050020 Almaty, Kazakhstan \\
      $^3$ Energetic Cosmos Laboratory, Nazarbayev University, 53 Kabanbay Batyr ave., 010000 Astana, Kazakhstan
      }

\maketitle

\begin{abstract}

We explore the dynamics of stellar discs in the close vicinity of a supermassive black hole (SMBH) by means of direct $N$-body simulations. We show that an isolated nuclear stellar disc exhibits anisotropic mass segregation meaning that massive stars settle to lower orbital inclinations and more circular orbits than the light stars. However, in systems in which the stellar disc is embedded in a much more massive isotropic stellar cluster, anisotropic mass segregation tends to be suppressed. In both cases, an initially thin stellar disc becomes thicker, especially in the inner parts due to the fluctuating anisotropy in the spherical component. We find that vector resonant relaxation is quenched in the disc by nodal precession, but it is still the most efficient relaxation process around SMBHs of mass $10^6\msun$ and above. Two body relaxation may dominate for less massive SMBHs found in dwarf galaxies. Stellar discs embedded in massive isotropic stellar clusters ultimately tend to become isotropic on the local two-body relaxation time-scale. Our simulations show that the dynamics of young stars at the centre of the Milky Way is mostly driven by vector resonant relaxation leading to an anticorrelation between the scatter of orbital inclinations and distance from the SMBH. If the $S$-stars formed in a disc less than 10 Myr ago, they may coexist with a cusp of stellar mass black holes or an intermediate mass black hole with mass up to $1000\msun$ to reproduce the observed scatter of angular momenta.

\end{abstract}


\section{Introduction}\label{sec:INTRO}

More than two decades of repeated monitoring of stellar orbits in the Galactic centre revealed the presence of a compact massive object that coincides with the radio source SgrA* \citep{Ghez2000, Gillessen2009, Genzel2010, Gillessen2017}. The high mass ($M \simeq 4\times10^6 \msun$) and compact size ($R < 10 ^{-6} \pc$) suggest that the object is a supermassive black hole (SMBH) (see \citealt{Eckart2017} for a discussion). The SMBH is surrounded by a dense cluster of stars, most of which are old (> 5 Gyr old), but some stars are very young (< 10 Myr old). The majority of young and massive stars are distributed in a disc-like structure as seen from their angular momentum vector directions \citep{LevinBeloborodov2003, PaumardEtAl2006, Bartko2009, Yelda2014,vonFellenberg2022}. This kinematic structure is called the clockwise stellar disc and is located between 0.04 and 0.5 pc \citep{LevinBeloborodov2003}. Another distinct kinematic structure is the $S$-star cluster: a cluster of young massive stars located within the inner arcsecond (0.04 pc) from the SMBH. Detailed spectroscopic studies of the $S$-stars indicate their ages are comparable with those of the clockwise stellar disc suggesting the same origin for both systems \citep{Habibi2017}. Recent observations suggest that the $S$-star cluster is likely to be arranged in two orthogonal discs \citep{Ali2020, Peissker2020} which may be identified from the distributions of the position angles of the semimajor axes of the sky-projected orbits \citep{Ali2020}\footnote{Note that the existence of two orthogonal discs in $S$-stars is debated \citep{vonFellenberg2022}}.  

The Milky Way galaxy is not the only galaxy that features a stellar disc. At the centre of the Andromeda galaxy, two distinct brightness peaks are observed \citep{Lauer1993} which may be explained by the so-called eccentric nuclear disc \citep{Tremaine1995} where orbits of stars have aligned arguments of periapsides. Observations of nuclear star clusters in nearby edge-on galaxies suggest that some of them host stellar discs associated with multiple stellar populations \citep{Seth2006, Seth2008}. Therefore, the coexistence of the nuclear star clusters with SMBHs and stellar discs appear to be common in the universe motivating studies of these systems. The main focus of this paper is the nuclear stellar disc of the Milky Way, but we also discuss stellar discs in nuclei of dwarf galaxies.

The interaction between a young stellar disc and the old spherical cluster may be described by secular processes that take place on time-scales significantly shorter than two-body relaxation. Due to the finite number of stars even a spherical cluster exhibits a fluctuating stochastic anisotropy which generates a strong net gravitational torque on stellar orbits, giving rise to rapid diffusion of orbital angular momenta in a process called resonant relaxation \citep{Rauch1996, Hopman&Alexander2006, Eilon2009, Kocsis2011, Kocsis2015,Giral2020}. In near-Kepler potentials in which the orbital time is much shorter than the apsidal precession time, the dynamics of stars can be represented as the interaction of quasi-stationary elliptical wires exerting mutual gravitational torques. In this case the individual orbital energies are approximately conserved, but the torques change both the magnitudes and the directions of the angular momentum vectors due to \textit{scalar} resonant relaxation (SRR). In non-Keplerian spherical mean-field potential, which arises in the Galactic centre due to the extended stellar mass distribution and/or general relativistic precession, the elliptical orbits are not closed, but experience rapid apsidal precession. For these systems the dynamical relaxation of orbital parameters is further accelerated by the coherent torques between $N$ rings or annuli covered by the individual stellar orbits. This reorients the angular momentum vector directions even more rapidly in a process called \textit{vector} resonant relaxation (VRR) while both orbital energy and angular momentum magnitude are nearly conserved \citep{Rauch1996}.

Theoretical studies of VRR benefit from the Hamiltonian formalism where the Hamiltonian represents the gravitational energy from the stellar potential excluding the Keplerian orbital energy around the SMBH \citep{Kocsis2015}. This may be achieved by orbit-averaging over the precession time-scale. The final equilibrium state may be found by means of mean field theory, the Monte Carlo Markov Chain method, kinetic theory, or by integrating Hamilton's equations of motion in time using orbit-averaged $N$-ring or direct $N$-body simulations.
First, using the mean field approach, the distribution function of the angular momentum vector directions can be found by maximising the entropy of the system using calculus of variations \citep{Roupas2017, Takacs2018, Magnan2021}. The equations have been solved analytically in the idealised case where all stars have identical masses, semi-major axes and eccentricities. \citet{Roupas2017} and \citet{Takacs2018} found  that  the stellar discs may represent statistical equilibrium structures. Moreover, depending on the total energy and angular momentum the system exhibits a phase transition between disc and spherical phases showing an analogy with liquid crystals. Recently these models were generalised by \citet{Magnan2021} to include the mass spectrum of stars showing that massive stars tend to arrange in thinner discs than light stars in a process called vertical mass segregation. This confirms the original expectation of \citet{Rauch1996}.

A similar conclusion was reached earlier using the Markov Chan Monte Carlo (MCMC) method. \citet{Szolgyen2018} showed that for a particular anisotropic initial condition the massive stars in the cluster form a disc. The study was recently extended by \citet{Mathe2022} where the authors explored the VRR equilibrium for a range of initial configurations in energy -- angular momentum space. Both of these studies included orbit-averaged interactions but did not consider the diffusion arising from two-body encounters. They found that massive objects form discs even in cases where the initial level of anisotropy is only a few percent.

Mass segregation may also occur in the eccentricity distribution, but in this case driven by SRR. Scalar resonant relaxation is the dominant process to randomise the eccentricities of the $S$-stars in the Galactic centre \citep{Perets2009b}. \citet{Fouvry2018} showed that massive stars tend to become more circular than light stars in discrete quasi-Keplerian axisymmetric discs. In spherically symmetric systems, mass segregation in eccentricity may take place in both directions: the orbits of massive stars become more circular and light stars become more eccentric or vice versa depending on the total energy of the system \citep{Gruzinov2020}.

The time-evolution of the system towards VRR equilibrium may be described by kinetic theory solving the Boltzmann equation. This approach has been used to elucidate SRR \citep{Bar-Or2018} and VRR processes \citep{Fouvry2019}. 

The time-evolution leading to mass-dependent anisotropy was demonstrated in a set of direct $N$-body and $N$-ring simulations featuring a stellar disc, an intermediate mass black hole (IMBH) and a spherically symmetric cluster of stars (implemented as an external potential) with a SMBH. \citet{Szolgyen2021} showed that the orbit of the IMBH aligns rapidly with the disc of stars within 3-10 Myr (depending on the IMBH mass and the initial inclination angle) and the IMBH eccentricity decreases rapidly due to VRR and SRR by the effect called resonant dynamical friction. This work featured direct integration of two-body encounters between the SMBH, IMBH, and the stars in the disc, but neglected the two-body interactions between stars in the disc and in the spherical cluster and deviations from spherical symmetry.

Mass segregation effects in the vicinity of a massive black hole were originally described in the context of two-body relaxation in isotropic spherically symmetric stellar systems \citep{Bahcall1977} which were later confirmed by direct $N$-body simulations \citep{Preto&Seoane2010, Panamarev2019}. For an isotropic system two-body relaxation is much slower than VRR by the ratio of the central mass to the individual stellar mass times $N^{1/2}$, i.e. $M_{\rm SMBH}/(N^{1/2}m)$, where $N$ is the number of stars. It drives mass segregation slowly both in semi-major axes and, as shown by \citet{Mikhaloff2017}, it leads to mass segregation in orbital inclinations and eccentricities in isolated stellar discs. Recently, $N$-body modelling of \citet{Foote2020} demonstrated vertical and eccentric mass segregation in eccentric nuclear discs. It was not clear from this study whether these effects were caused by two-body or resonant relaxation or both. Anisotropic mass segregation was also observed in direct $N$-body simulations of rotating globular clusters \citep{Szolgyen2019}, where VRR dominates over two-body relaxation for $N\gg 10^4$ \citep{Meiron2019}.

\citet{Perets2018} showed that the collective effect of stars in a spherical distribution (in their case a cusp of stellar black holes) may lead to the formation of clumps, warps and spiral arms in the stellar disc. They compared results from direct $N$-body simulations of isolated stellar discs, stellar discs embedded in a smooth potential, a hybrid self consistent field modelling of disc -- sphere interactions \citep{Meiron2014} and direct $N$-body integration of the whole system. While isolated discs and discs embedded in a smooth potential showed steady increase in disc thickness, both hybrid and direct $N$-body models led to the formation of clumps, warps and spiral arms. The qualitative agreement between hybrid and direct models suggests that these effects may be caused by resonant relaxation.

\citet{Mastrobuono-Battisti2019} used direct $N$-body simulations to study the co-evolution of multiple stellar discs embedded in an analytic stellar cusp and a discrete population of stellar black holes. By introducing a new disc every 100 Myr, they found that the discs evolve towards a uniform distribution in orbital inclinations, but at the end of their simulations (500 Myr) each of the discs showed different morphologies and kinematics. 

\citet{Kocsis2015} and \citet{Giral2020} showed that the fluctuating anisotropy of a spherical distribution leads to diffusion in angular momentum direction space in a nearly spherical system due to VRR. Thus, as long as the gravitational interaction between disc particles may be neglected, a spherical distribution drives the disruption of a stellar disc. Furthermore, two-body relaxation may further accelerate rapid diffusion, rapidly increasing the thickness of an initially very thin disc \citep{Cuadra2008}. In the opposite limit of a strongly self-interacting thin stellar disc with no two-body relaxation, the disc acts as a coupled system of harmonic oscillators, counteracting the external torques such that the disc remains intact and exhibits normal mode oscillations \citep{Kocsis2011}. In this paper we aim to study the interaction of a nuclear stellar disc with a spherical nuclear star cluster around a central massive black hole self-consistently by means of direct $N$-body simulations.  We improve the physical realism and particle number resolution over previous direct $N$-body models to understand if stellar discs or black hole discs may be long lived in nuclear star clusters.

The paper is organised as follows. In Sec. \ref{sec:time-scales} we review the Galactic centre time-scales. In Sec.~\ref{sec:simulations} we describe the initial setup for our numerical models. Sec.~\ref{sec:disk-evol} is devoted to the analysis of isolated stellar discs without a spherical stellar population, and Sec.~\ref{sec:disk-sphere-evol} to the effects caused by the dynamical interaction with the sphere. In Sec.~\ref{sec:s-stars} we apply our findings to compare with the observed population of $S$-stars and, finally, we summarise the paper in Sec.~\ref{sec:SUM}.

\section{The time-scales}
\label{sec:time-scales}

\begin{figure*}
	\begin{center}
		\includegraphics[width=\linewidth]{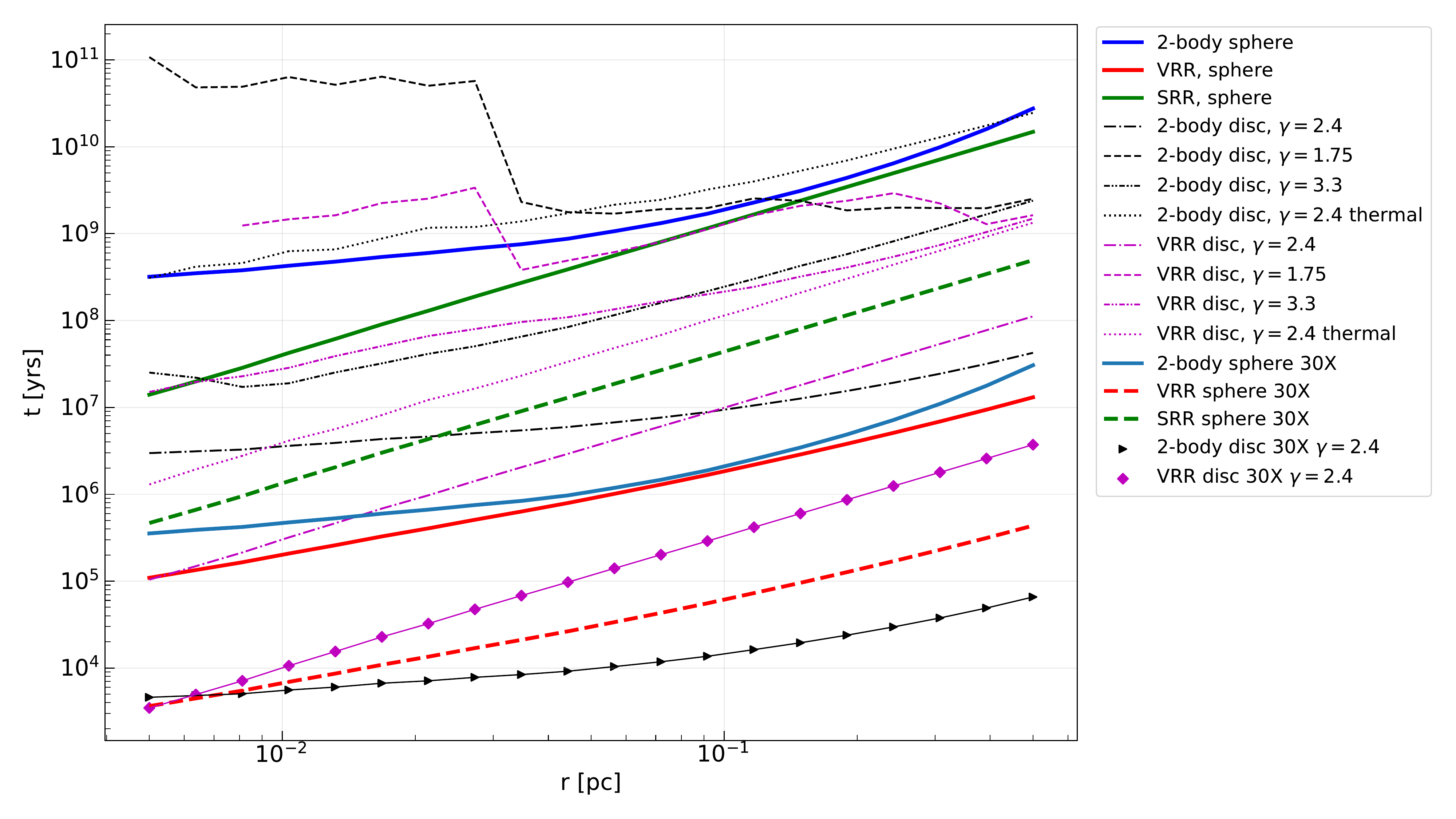} \\ 
		\par\end{center}
	\caption{The time-scales of dynamical processes as a function of distance from the SMBH in our simulations of the Galactic centre. Computed from analytical expressions presented in Sec.~\ref{sec:time-scales} using the data from the initial conditions of the models described in Sec.~\ref{sec:simulations}. All thick lines show the time-scales related to the interaction of the stellar disc with a spherical component with a 3D density distribution $\rho\propto r^{-1.75}$.  All thin lines show the relaxation time-scales within the disc neglecting contribution from a spherical component. Black lines illustrate two-body relaxation within the discs, purple lines show VRR within the discs. Different line styles correspond to density distributions of the stellar discs with corresponding power-law density slope according to the legend.} 
	\label{fig:time}
\end{figure*}

In this section we review the relaxation processes in galactic nuclei and the associated time-scales similar to \citet{Kocsis2011} and \citet{Rauch1996}.

\paragraph*{Two-body relaxation.}
Two-body relaxation arises from the fluctuating force acting on a subject star over the orbital period. As the total impulses received by a star over the orbital period are uncorrelated, the relaxation rate occurs in a random-walk fashion and is often called non-coherent relaxation. The two-body relaxation changes both the energy and the corresponding angular momentum at the rate (see e.g. \citealt{Rauch1996} or \citealt{BinneyTremaine2008}):
\begin{equation}
\label{eq:E-2b}
\frac{\Delta E}{E} =\alpha \frac{m_2 N^{1/2}}{M_\mathrm{bh}}\left(\frac{t}{t_\mathrm{orb}}\right)^{1/2},\quad
\frac{\Delta L}{L} = \beta \frac{m_2 N^{1/2}}{M_\mathrm{bh}}\left(\frac{t}{t_\mathrm{orb}}\right)^{1/2},
\end {equation}
where $N$ is the total number of stars, $M_\mathrm{bh}$ is the mass of the central massive black hole, $E \sim 2GM_\mathrm{bh}/R $ is the Keplerian energy, $m_2 = \langle m^2\rangle/\langle m\rangle$ is the effective mass and $\alpha\sim\beta\sim (\ln \Lambda)^{1/2}$ within factors of order unity where $\ln\Lambda\simeq\ln(M_\mathrm{bh}/m)$ is the Coulomb logarithm, $m$ is the stellar mass and $t_\mathrm{orb}$ is the orbital period.

The two-body relaxation time-scale for a spherical stellar system with a central massive black hole can be computed by \citep{BinneyTremaine2008}:
\begin {equation}
\label{eq:t-2body}
t_\mathrm{relax} = 0.34\frac{\sigma^3(r)}{G^2\rho(r) m_2\ln\Lambda} = \frac{M_{\rm bh}^2}{\beta^2 m_2^2 N} t_{\rm orb},
\end {equation}
where $\sigma$ is the one-dimensional velocity dispersion, $\rho$ is the stellar density.

\paragraph*{Scalar resonant relaxation.}
Contrary to two-body relaxation, SRR occurs in a coherent way over the apsidal precession time-scale. In near-Kepler potentials, the orbit-averaged interaction may be approximated as elliptic wires exerting mutual torques. In this case the Keplerian energy is conserved, but both the magnitude and the direction of angular momentum vectors $L$ are changed at the following rate:
\begin {equation}
\label{eq:L-srr}
\frac{\Delta L}{L_{\rm c}} = \eta_s \frac{m_2 N^{1/2}}{M_\mathrm{bh}}\left(\frac{t_\mathrm{prec} t}{t_{\rm orb}^2}\right)^{1/2},
\end {equation}
where $L_{\rm c}=L/\sqrt{1-e^2}$, $\eta_\mathrm{s}$ is a dimensionless coefficient of order unity and $t_\mathrm{prec}$ is the apsidal precession time. The total relaxation rate occurs in a random walk fashion with the apsidal precession time being the step size (duration of the coherent phase). The long duration of the step size compared to the orbital period makes this process more efficient than two-body relaxation in near-Kepler potentials where $t_\mathrm{prec} \gg t_\mathrm{orb}$.

The SRR time in a spherical stellar system can be found by:
\begin {equation}
\label{eq:t-srr}
t_\mathrm{rr,s} = \frac{4\mathrm{\pi} |\omega|}{\beta_s^2\Omega^2}\frac{M^2_\mathrm{bh}}{M(r)m_2},
\end {equation}
where $\omega=2\pi/t_{\rm prec}$ is the apsidal precession rate (sum of Newtonian and relativistic), $\Omega=2\pi/t_{\rm orb}$ is the orbital frequency and $\beta_s$ is a dimensionless coefficient estimated by \citet{Eilon2009} to be $1.05 \pm 0.02$. 

\paragraph*{Vector resonant relaxation.}
In spherical potentials where the precession time is short, the stellar orbits may be approximated as annuli that exert mutual torques. In this case the torques change the direction of orbital angular momentum vectors at the rate:
\begin {equation}
\label{eq:L-vrr}
\Delta \boldsymbol{L}/L_{\rm c} = \eta_\nu \frac{m_2 N^{1/2}}{M_\mathrm{bh}}\left(\frac{t}{t_\mathrm{orb}}\right)^{1/2} + \beta_\nu \frac{m_2 N^{1/2}}{M_\mathrm{bh}}\frac{t}{t_\mathrm{orb}},
\end {equation}
where $\eta_\nu$ is a dimensionless coefficient which corresponds to the contribution of two-body relaxation and SRR, and the term with $\beta_\nu = 1.83\pm 0.03$ represents the contribution from the coherent phase of VRR (linear with $t/t_\mathrm{orb})$ \citep{Eilon2009}. \citet{Kocsis2015} found that VRR is slower by a factor 3 due to rapid apsidal precession consistent with earlier work \citep{Rauch1996}. It is expected that VRR may be the most efficient way to randomise the stellar orbital inclinations as the step size of the coherent phase is the largest among all relaxation processes.

For a spherical stellar system, the VRR time is \citep{Eilon2009}:
\begin {equation}
\label{eq:t-vrr}
t_\mathrm{rr,v} = \frac{M_\mathrm{bh}}{\sqrt{M(r)m_2}}\frac{t_{\rm orb}}{\beta^2_\nu}.
\end {equation}
\citet{Kocsis2015} found that $m_2$ is replaced by the RMS mass for VRR.

\paragraph*{Two-body relaxation in a stellar disc.}
Two-body relaxation time-scale for a stellar disc can be computed by \citep{Stewart2000}:
\begin {equation}
\label{eq:t-2b-disk}
t_\mathrm{rx, disc} = \frac{\left<e^2\right>^2}{4.5\Omega} \frac{M^2_{bh}}{m_2\Sigma r^2\ln{\Lambda}},
\end {equation}
where $\Sigma$ is the surface density of the disc, $\Lambda\simeq\left<e^2\right>^{3/2}M_\mathrm{bh}/m$. The formula assumes $\left<i^2\right>^{1/2}\simeq0.5\left<e^2\right>^{1/2}$

\paragraph*{Vector resonant relaxation in a stellar disc.}
VRR may also occur in stellar discs. Since stars exert torques from the disc plane leading to precession in the line of nodes at the rate \citep{Kocsis2011}:
\begin {equation} \label{eq:nu-nodal}
\nu \simeq \frac{\Omega}{\left<i^2\right>^{1/2}} \frac{M_\mathrm{disc}}{M_\mathrm{bh}},
\end {equation}
the nodal precession will limit the step size for the coherent phase of VRR. To compute VRR in a stellar disc, we replace the apsidal precession rate in Eq.~\ref{eq:t-srr} by the nodal precession rate and $M_\mathrm{disc}$ by $M(r)$:
\begin {equation}
t_\mathrm{vrr, disc} \simeq  \frac{4\pi}{\Omega\left<i^2\right>^{1/2}}\frac{M_\mathrm{bh}}{m_2}.
\end {equation}
This expression shows relaxation of the angular momentum vectors which in this case is dominated by relaxation in azimuthal components driven by the nodal precession (as shown in Sec.~2 of \citealt{Kocsis2011}). Note that VRR in the vertical direction may be much slower due to kinetic blocking \citep{Fouvry2019a}. Furthermore, $t_\mathrm{vrr, disc}$ estimates the timescale for the relaxation of a disc by neglecting the fluctuating torques from the spherical component of the stellar distribution. 

We refer to \citet{Tremaine1998} and \citet{Fouvry2018} for the discussion and analysis of SRR in discs.

For the relaxation processes that occur much faster than two-body relaxation, it is often useful to compare the time-scales with respect to the secular time, defined as:
\begin {equation}
\label{eq:tsec}
t_\mathrm{sec} = \frac{M_\mathrm{bh}}{M_\mathrm{tot}}P_\mathrm{inner},
\end {equation}
where $P_\mathrm{inner}$ is the orbital period of the innermost star (in our models determined by the inner edge of the stellar disc) and $M_\mathrm{tot}$ is the total stellar mass of the system. This time-scale sets the shortest apsidal precession time. 

Fig.~\ref{fig:time} shows the time-scales described above applied to the Galactic centre using data from our simulations (see Sec.~\ref{sec:simulations}). The spherical component corresponds to the Bahcall-Wolf cusp \citep{Bahcall1976} while stellar discs feature various distributions of 3D densities and orbital parameters adopted in our simulations as described in the following section. The figure compares the time-scales of dynamical processes within the sphere (thick lines) and within the discs (thin lines). As we see, VRR within the sphere (thick red line) is the fastest process followed by VRR in discs (although for some disc models 2-body relaxation within the disc is comparable in some regions; see purple and black lines). On the other hand, if the total mass of the whole stellar system is increased by a factor of 30 (labelled as 30X in the legend), while keeping the same number of particles, 2-body relaxation within the disc becomes the fastest process (see the section below for a motivation on the 30X models). 

Note that the time-scales presented in the Figure~\ref{fig:time} (and the equivalent analytical expressions) are derived either neglecting the contribution from the disc (time-scales within the sphere) or from the sphere (time-scales within the discs), but in reality the dynamics of a stellar disc embedded in a sphere may be shaped by the contribution from both the disc and the sphere. 
The torque acting on a test particle in the presence of an isotropic cluster due to the fluctuating stochastic anisotropy is of order \citep{Kocsis2015}
\begin{equation}
    \dot{L}_{\rm sphere} \sim \beta_\nu \frac{N_{\rm sphere}^{1/2} m_{\rm rms,sphere}}{M_{\rm bh}} \frac{L_{\rm c}}{t_{\rm orb}}\,,
\end{equation}
while a stellar disc drives nodal precession at the rate of order
\begin{equation}
    \dot{L}_{\rm disc} \sim \frac{N_{\rm disc} m_{\rm av, disc}} {M_{\rm bh}} \frac{L_{\rm c}}{t_{\rm orb}}\,.
\end{equation}
Here $m_{\rm rms,sphere}=\langle m^2\rangle^{1/2}$ and $m_{\rm av, disc}=\langle m\rangle$ for objects in the spherical cluster and the disc, respectively. Thus, the effect of the disc dominates over the sphere if $N_{\rm disc} m_{\rm av, disc} \gg N_{\rm sphere}^{1/2} m_{\rm rms, sphere}$ and the disc exhibits normal mode oscillations \citep{Kocsis2011}, and in the opposite limit the disc dissolves on the $t_{\rm rr,v}$ VRR timescale due to the sphere \citep{Kocsis2015,Giral2020}.
To explore the dynamics and the dominant relaxation process for different systems, we perform direct $N$-body simulations of stellar discs embedded in a spherical cusp of stars in the intermediate regime where $N_{\rm disc} m_{\rm av, disc}$ and $N_{\rm sphere}^{1/2} m_{\rm rms, sphere}$ are comparable as we describe in the following section.

\section{Simulations}
\label{sec:simulations}

We adopt the following system of units for all the models:
\begin{equation}\label{eq:units}
G=M_\mathrm{bh}=R_\mathrm{out}=1,
\end{equation}
where $G$ is the gravitational constant, $M_\mathrm{bh}$ is the initial SMBH mass, and $R_\mathrm{out}$ is the initial outer radius of the stellar system which is defined as the orbital semi-major axis of the outermost star in the system.  When converting to physical units we typically assume $R_\mathrm{out}=0.5$ pc, $M_\mathrm{bh}=4\times10^6\msun$ unless indicated otherwise, and in some cases we adopt $R_\mathrm{out}=1$ pc, $M_\mathrm{bh}=1.3\times10^5\msun$.

\subsection{The code}
\label{subsec:code}

We use a modified direct $N$-body code \textsc{$\varphi$-grape} \citep{HarfstEtAl2007} that uses 4-th order Hermite integration method \citep{Makino1991, Makino1992, Aarseth2003} to solve the equation of motion. The code was originally designed for the GRAPE cards and now utilises an emulation library to run on modern GPUs \citep{Nitadori2008}. The modified version of the original code includes the gravitational interaction with the massive central object implemented as a fixed external point-mass potential and the accretion of stars onto the central object  
\citep{JYM2012, LLBCS2012, ZBS2014}. 
The equation of motion is:
\begin{equation}\label{eq:motion}
\ddot{\boldsymbol{r}_{i}}
=-\sum_{i\not= j}\frac{Gm_\mathrm{j}\boldsymbol r_{ij}}{(r_{ij}^2 + \epsilon_\mathrm{ss}^2)^{3/2}} -
\frac{GM_\mathrm{bh}\boldsymbol r_{i}}{r_{i}^3}, 
\end{equation}
where $\boldsymbol r_{ij}=\boldsymbol r_{i}-\boldsymbol r_{j}$ with $\boldsymbol r_{i}$,
$\boldsymbol r_{j}$ the positions of stars $i$ and $j$, respectively, $\epsilon_\mathrm{ss} = 1.0\times10^{-4}$ is the stellar softening parameter. The value for the softening between stars is chosen to be small enough to resolve relevant close encounters but large enough to prevent formation of the compact binary systems. Lower value for the softening may result in a larger number of very close encounters between stars, but they are rare and are not relevant on the resonant relaxation time-scales which are the main focus of this work. 

The central massive black hole can grow in mass by consumption of stars. The criterion for the accretion is the instantaneous distance to the star is less than the accretion radius which was set to be equal to the tidal disruption radius of a $2R_\odot$ star by a $4\times10^6\msun$ black hole. After the accretion event the total mass of the star is instantaneously added to the mass of the SMBH and the star is removed from the simulation \citep{JYM2012, LLBCS2012, ZBS2014}. The accretion radius sets the innermost resolution of the simulations and, thus, allows not to soften the interaction between stars and the SMBH \citep{KCMB2018}. 

The accuracy of the simulations is controlled by the time-step factor $\eta$ \citep{Aarseth1985, Makino1992}. We choose $\eta = 0.01$ as a compromise between the accuracy and the computing time. To ensure that $\eta = 0.01$ is the optimal choice, one can measure the total energy exchange between particles caused by two-body relaxation over the apsidal precession time and compare it to the total absolute energy error of the system over the same period of time. For all of our models, the ratio of the absolute energy error over the total energy exchange between particles does not exceed $10^{-5}$ over the apsidal precession time for a given particle ensuring that $\eta = 0.01$ is the optimal choice. The total relative energy error at the end of the simulations is of order $\Delta E = \frac{E - E_0}{E_0} \approx 10^{-4}$, the total angular momentum error is of order $\Delta\mathbf{L} = \frac{|\mathbf{L} - \mathbf{L}_0|}{|\mathbf{L}_0|} \approx 10^{-3}$. Reducing the value for $\eta$ improves the error tolerance, but slows down the computations and qualitatively shows the same results. 

\subsection{Initial conditions}
\label{subsec:IC}

We study the gravitational interaction of a galactic nucleus with three components: a central massive black hole, a spherical cluster of old stars and a population of stars resembling a disc. 

We run  one-to-one simulations meaning that one particle in the simulation represents one realistic star. This can be achieved by modelling a system of $10^5$ particles with an average particle mass of $10^{-6}\,M_{\rm bh}$. Using a top-heavy initial mass function (IMF, Eq.~\ref{eq:kroupa-sphere} below) and applying the parameters to the Milky Way Galaxy centre gives the total stellar mass $M_\mathrm{tot} \simeq 2\times 10^5 \msun$ for the inner $0.5$\,pc. This value is comparable to the total stellar mass inferred from observations: \citet{Schodel2018} find $M \simeq 1.3 \times 10^4 \msun$ within 0.1 pc and  $M \simeq 1.0 \times 10^6 \msun$ within 1 pc.\footnote{Note that these estimates do not include stellar remnants meaning that the actual enclosed mass within the regions may be higher.} The most recent estimates based on interferometric astrometry indicate that the total extended mass within 0.1 pc does not exceed $M\simeq 10^5 \msun$ \citep{Gravity2022}. 

We generate the initial positions and velocities for the spherical stellar system to follow Keplerian orbits with spatial density distribution resembling a Bahcall-Wolf cusp with $\rho\propto r^{-7/4}$ where $r$ is the distance from the SMBH \citep{Bahcall1976}. The distribution of orbital parameters for the spherical cluster is the same in all our models while we vary the spatial density distribution and orbital parameters for the disc stars as described in Sec.~\ref{subsec:IC}. In all the models we keep the stellar disc embedded in a spherical component. 

To model the mass spectrum of stars, we adopt the \citet{Kroupa2001} top-heavy IMF for the sphere:
\begin{equation}\label{eq:kroupa-sphere}
\frac{dN}{dm} \propto m^{-\alpha},\quad 
\alpha_{\rm sphere}=
\left\{
\begin{array}{cl}
      1.3,&   {\rm if~} 0.08\msun \leq m < 0.5\msun\\
      2.3,&   {\rm if~} 0.5\msun \leq m < 1.0\msun \\
      1.5,&   {\rm if~} m \geq 1.0\msun
\end{array}
\right.
\end{equation}
The top-heavy IMF is motivated by the expected mass segregation in  galactic nuclei (see e.g. \citealt{Panamarev2019}), and the observed stellar mass function in the Galactic centre following $m^{-1.7 \pm 0.2}$ \citep{Lu2013}. After the IMF is generated we use the stellar evolution code (SSE, \citealt{Hurley2000}) to evolve the whole system up to 1 Gyr and use stellar masses at 1 Gyr as the initial mass distribution for both disc and spherical components. This allows us to ignore the mass loss due to the stellar evolution in the code during the dynamical evolution.

\begin{table}
\caption{List of simulations with a nuclear stellar disc and sphere}
\label{tab:runs}
\begin{center}
\begin{tabular}{lll}
\hline\hline
Fiducial models:\\
\hline
Mass factor & Initial orbital & Disc 3D density    \\
 & parameters &    \\
\hline
1; 10; 30  & stardisc & 1.75 \\
   &          & 2.4 \\
   &    & 3.3 \\
\hline   
1; 10; 30   & stardisc-random & 2.4   \\
\hline   
1; 10; 30   & thermal & 2.4 \\
\hline\hline
Additional \\models:\\
\hline
$N_\mathrm{d}$ & $N_\mathrm{s}$ &  $M_\mathrm{d}/M_\mathrm{s}$ \\
\hline
$10^4$ & $10^5$ & 0.14 \\
$10^3$ & $10^5$ &  0.04 (massive disc)\\
$5\times10^4$ & $5\times10^4$ & 1.0  \\
$9\times10^4$ & $10^4$ & 8.8  \\ 
\hline
\end{tabular}
\end{center}
\textbf{Notes.} List of models with different initial conditions for stellar discs. The default number of stars in the disc and the sphere are $N_\mathrm{d}=10^3$ and $N_\mathrm{s}=10^5$, respectively; the radial number density profile exponent of the sphere and the disc are -1.75 and $\gamma=-2.4$. For the \textit{stardisc} initial conditions we also adopted two additional $\gamma$ values as shown. For each of these main models, we adopted three different mass factors to scale the stellar mass distribution as shown  to accelerate the code (see text). In total, for the main models we have 9 \textit{stardisc} models, 3 \textit{stardisc-random} and 3 \textit{thermal} models. For the additional models the mass factor is 30, the disc radial density profile slope is $\gamma=2.4$ and initial orbital parameters are \textit{thermal}.
\end{table}

We use a slightly shallower slope for the heavier masses but keep the same break points to generate the IMF for the stellar disc motivated by observations \citep{Bartko2010}\footnote{Galactic center observations suggest an even more top-heavy profile $dN/dm\propto m^{-0.45\pm0.3}$ \citep{Bartko2010}}:
\begin{equation}\label{eq:kroupa-disk}
\alpha_{\rm disc}=
\left\{
\begin{array}{cl}
      1.3,&   {\rm if~} 0.08\msun \leq m < 0.5\msun\\
      2.3,&   {\rm if~} 0.5\msun \leq m < 1.0\msun \\
      1.3,&   {\rm if~} m \geq 1.0\msun
\end{array}
\right.
\end{equation}

\begin{figure*}
	\begin{center}
		\includegraphics[width=\linewidth]{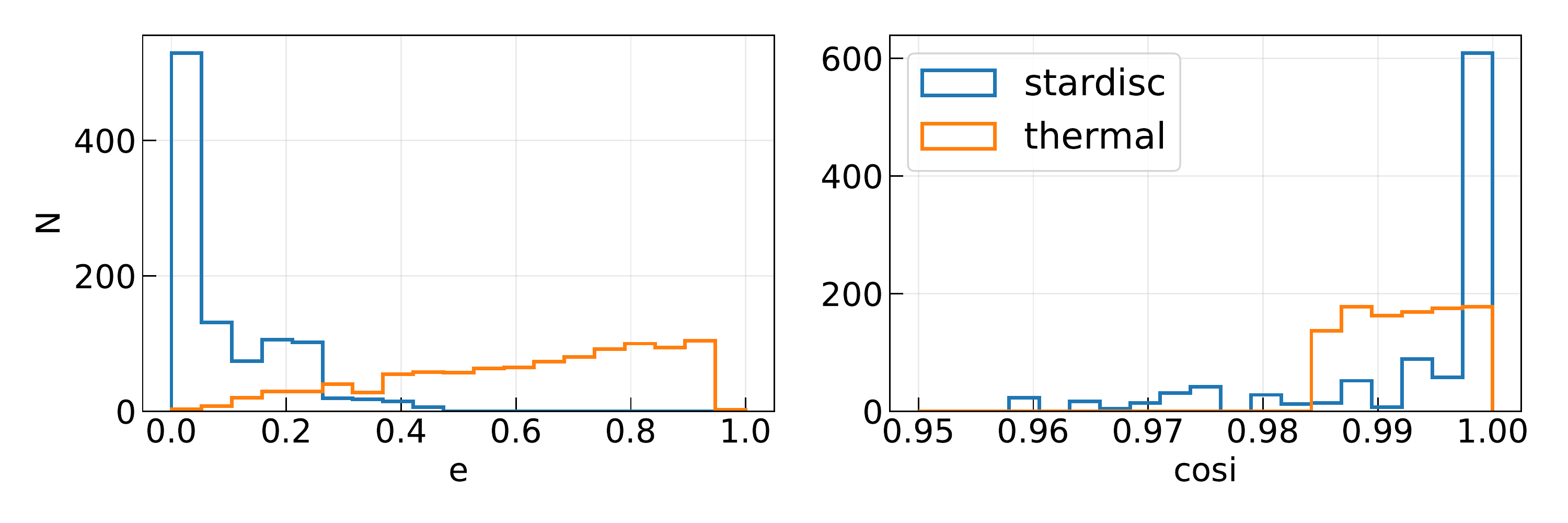} \\ 
		\par\end{center}
	\caption{\textit{Left panel:} Initial distribution of eccentricities for the stellar disc. Blue histogram shows the initial conditions originating from the stardisc simulations \citep{Panamarev2018} of active galactic nuclei while the orange histogram represents thermal eccentricity distribution. \textit{Right panel:} Distribution of cosines of inclination angles for the stellar disc. Blue shows the \textit{stardisc} initial conditions and orange line corresponds to the \textit{thermal} model: uniform distribution in cos(i) corresponding to angles between 0 and 10$^{\circ}$}
	\label{fig:ini_distrl}
\end{figure*}

\begin{figure*}
	\begin{center}
		\includegraphics[width=\linewidth]{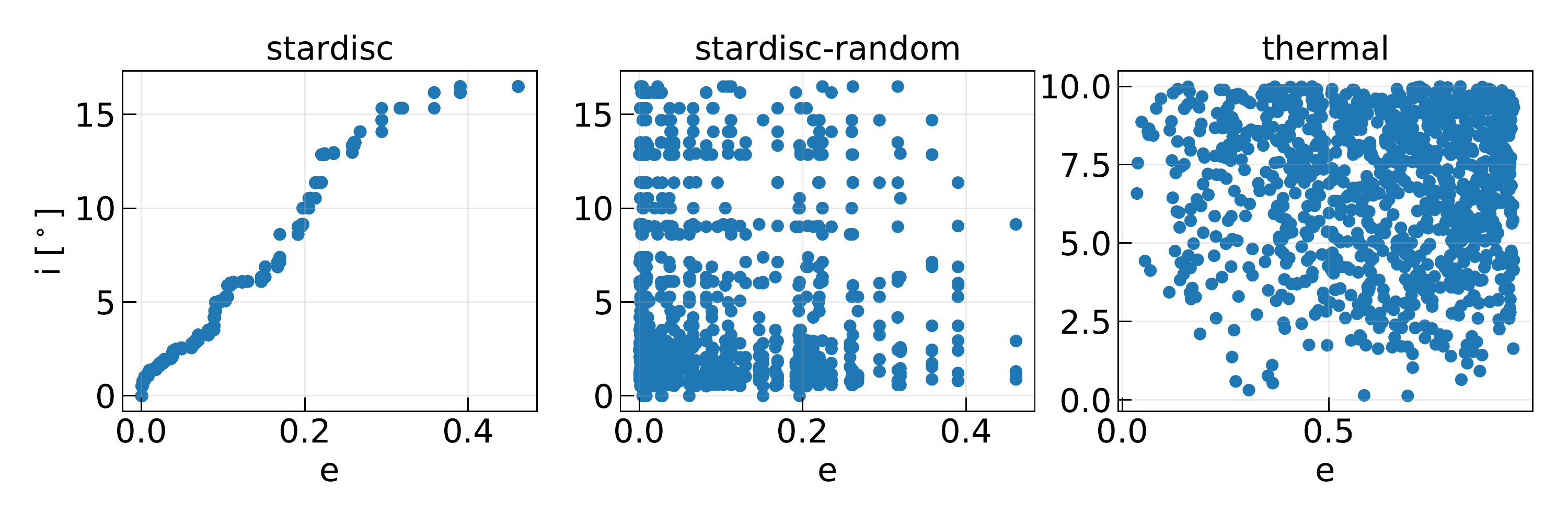} \\ 
		\par\end{center}
	\caption{Scatter plot of stellar disc eccentricities and inclination angles for the three types of initial conditions adopted. Left panel, labelled \textit{stardisc}, shows the relation between eccentricity and inclination angle arising from previous  stardisc simulations in AGNs \citep{Panamarev2018}. Middle panel, labelled \textit{stardisc-random}, shows the model in which the correlation is removed by independently assigning inclinations and eccentricities from the stardisc model, and the right panel shows the \textit{thermal} model (see text). }
	\label{fig:ini_rel}
\end{figure*}

We explore several models for the distribution of orbital parameters in the disc as summarised in Table~\ref{tab:runs}. We consider two main scenarios for the origin of the stellar disc. The first one is the formation of the disc due to the star -- disc interactions in an active galactic nucleus. \citet{Panamarev2018} showed that the gaseous accretion disc may capture stars from the surrounding star cluster with the captured stars following the disc-like shape resembling the shape of the underlying gaseous accretion disc (see also \citealt{Bartos2017}). The formed stellar disc is in steady state balanced by the accretion of stars onto the SMBH and capturing new stars by the accretion disc. To generate the initial positions and velocities, we take data from \citet{Panamarev2018} at 1 relaxation time (enough to form the steady state disc) and make statistical bootstrapping to increase the number of stars (in \citealt{Panamarev2018} the authors had to use the super-particle approach where 1 particle represented a group of stars). First, we convert positions and velocities to 6 Keplerian orbital parameters (this is a good approximation for orbits deep inside the influence radius of the SMBH), then generate a larger number of objects corresponding to the distribution function of orbital parameters, and finally, we convert the orbital parameters back to positions and velocities. This way we generate 1000 particles for our models from the original $\approx$100 particles taken from \citet{Panamarev2018}. We refer to the initial orbital parameters of the disc stars derived this way as the \textit{stardisc} initial conditions. Fig.~\ref{fig:ini_distrl} (blue lines in both panels) shows notable features: nearly circular orbits for most of the stars and low orbital inclinations. There is also a linear dependence of the orbital inclination, eccentricity and semi-major axis which resembles the outer warp of the stellar disc (see the left panel of the Fig.~\ref{fig:ini_rel} that shows the correlation between the inclination angles and eccentricities). 

As this type of initial conditions may seem specific to the underlying accretion disc model used in \citet{Panamarev2018}, we explored another family of the \textit{stardisc} initial conditions where we kept the same distributions of the orbital parameters as in Fig.~\ref{fig:ini_distrl}, but randomised the inclination -- eccentricity -- semi-major axis relation as shown in the middle panel of Fig.~\ref{fig:ini_rel}. We refer to these initial conditions as the \textit{stardisc-random} initial conditions. In the \textit{stardisc} initial condition models we vary the 3D density power-law slope for semi-major axes as described in Sec.~\ref{subsec:IC}. 

In addition to the \textit{stardisc} and \textit{stardisc-random} initial conditions, we also explore the case where the stellar disc follows a thermal eccentricity distribution, uniformly distributed orbital inclinations between $\cos{10^{\circ}}$ and $\cos{0^{\circ}}$, and a 3D power-law density slope for the semi-major axes $\rho \propto r^{-2.4}$ implying that $dN/da = a^{-0.4}$. Orange lines in both panels of Fig.~\ref{fig:ini_distrl} and the right panel of Fig.~\ref{fig:ini_rel} highlight the differences between the models. We refer to these initial conditions as \textit{thermal} initial conditions. The remaining Keplerian orbital elements, namely longitudes of the ascending nodes, arguments of periapsis and mean anomalies are drawn from a uniform distribution within the whole range of their allowed values.

We perform a set of simulations with $N_\mathrm{s} = 10^5$ total number of stars in the sphere, $N_\mathrm{d} = 10^3$ total number of stars in the disc and average mass ratio of $m_*/M_\mathrm{bh} = 5\times 10^{-7}$. Given the slightly different mass functions for the disc and for the sphere the total mass fraction of the disc is $M_\mathrm{d}/M_\mathrm{s} \simeq 0.015 $. 
To explore the effects of the initial orbital parameters distribution we use 3 sets of models: \textit{stardisc}, \textit{stardisc-random} and \textit{thermal}, as described above. For the \textit{stardisc} model we vary the power law slope for the 3D density distribution $\rho \propto r^{-\gamma}$ with $\gamma = 1.75$ to represent the standard Bahcall-Wolf cusp \citep{Bahcall1976}, $\gamma = 2.4$ to match the observed density distribution of the clockwise stellar disc in the Galactic centre \citep{Yelda2014} and $\gamma = 3.3$ -- the steepest density profile in our models which originates from the star -- disc simulations of \citet{Panamarev2018}, for other models we fix $\gamma=2.4$. This gives us 5 different models which are referred as $1X$ models. Due to the high numerical cost, these types of simulations can be advanced up to 5-10 Myr when applied to the Galactic centre corresponding to the observed age of the nuclear stellar disc and $S$-stars \citep{Habibi2017}.

To study long term evolution of the system, we increase the total stellar mass of the system by factors of 10 and 30 respectively while keeping the same number of particles. This gives 10 more models. We refer to these models as $10X$ and $30X$ models. As we saw in Sec.~\ref{sec:time-scales}, the dynamical time scales are reduced for a larger total stellar mass. Due to the fact that the scaling with mass is different for the resonant relaxation and for the two-body relaxation (see Eq.~\ref{eq:t-vrr} and Eq.~\ref{eq:t-2body}), we can study the contribution from these relaxation processes by comparing the $1X$, $10X$ and $30X$ models. Table~\ref{tab:runs} lists all the models and their parameters. 

The bottom part of the Table~\ref{tab:runs} lists several additional models that we simulated with the \textit{thermal} initial conditions and $\gamma=2.4$ disc density exponent. First we include additional variants of the $30X$ models, which are numerically the least expensive and allow us to explore the parameter space of the system. In particular, we run additional models with (i) a larger number of stars in the disc $N_\mathrm{d} = 10^4$; and (ii) with the same number of stars in the disc but increased total mass of the disc. Furthermore, we examine two additional models where the number of stars in the disc was equal to the number of stars in the sphere and where the number of stars in the disc was 90\% of the total number of particles with the total number of particles $N=10^5$ in both runs. 
In addition, in order to study the effect of the sphere on the dynamics of stars within the disc, we run the fiducial $1X$, $10X$ and $30X$ models without the sphere, with only a stellar disc of $N_\mathrm{d} = 10^3$ stars around the supermassive black hole. We refer to these models as the \textit{isolated disc} models.

\section{Dynamics of the isolated stellar discs}
\label{sec:disk-evol}

In this section we describe the evolution of isolated stellar discs rotating around a SMBH with 100\% of stars initially on prograde orbits and no spherical stellar component. As reference models we choose the \textit{thermal} models with the power-law density slope of the disc $\gamma=2.4$ and the mass factors 1, 10 and 30. We examine how the total stellar mass (with fixed number of particles) affects the dynamics of the relaxation processes. 

\begin{figure*}
	\begin{center}
		\includegraphics[width=\linewidth]{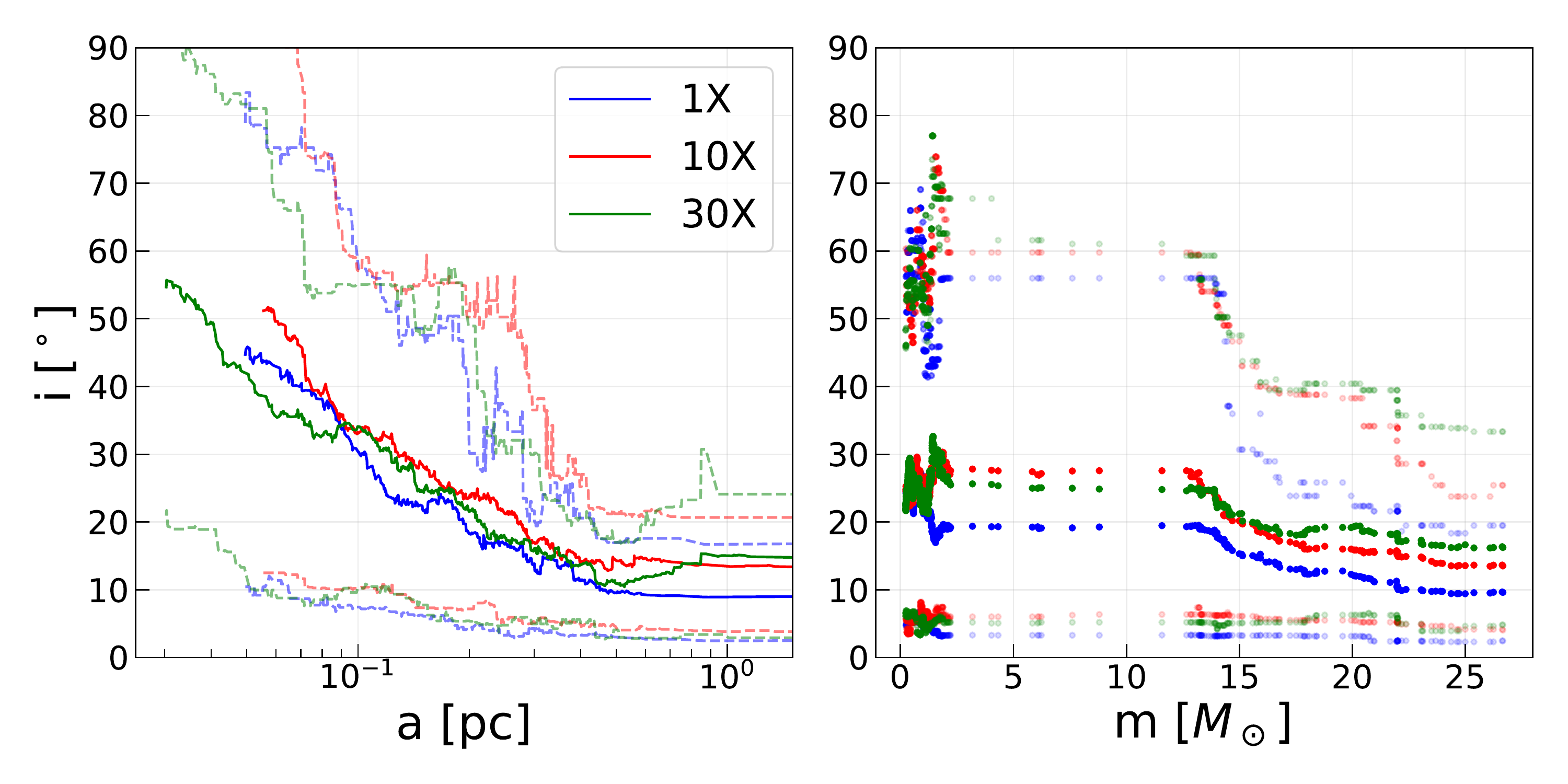} \\ 
		\par\end{center}
	\caption{Left panel: moving average of inclination angles in semi-major axes for the $thermal$ isolated disc models (see Table~\ref{tab:runs}) with mass factors according to the legend. Right panel: moving average of inclination angles in mass for the same models. Both panels correspond to the same time snapshot. Blue, red and green colours indicate 1X, 10X and 30X \textit{thermal} isolated disc models, respectively. Faded lines (points) of the same colour show 90 and 10\% quantiles. The time snapshots for each model are chosen such that average orbital inclinations match for the low mass stars (this corresponds to 56, 5 and 0.9 Myr for 1X, 10X and 30X models respectively; see dotted lines in Fig.~\ref{fig:rms-od}). For the 10X and 30X models (red and green lines in the right panel) we show m/10 and m/30 respectively so that the mass ranges overlap. The window for the moving averages was chosen to be 100 data points.}
	\label{fig:i-a-od}
\end{figure*}

\begin{figure*}
	\begin{center}
		\includegraphics[width=\linewidth]{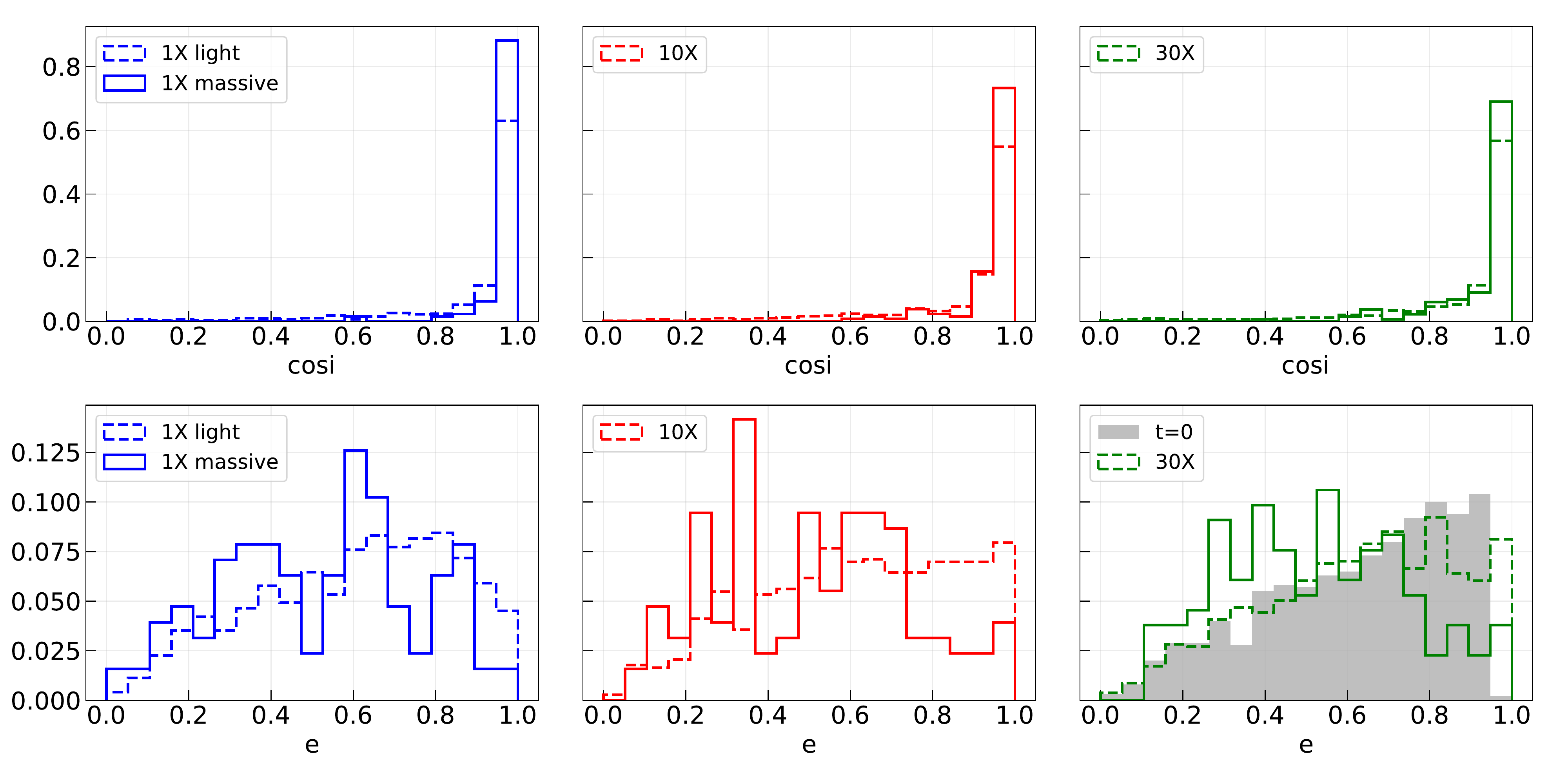} \\ 
		\par\end{center}
	\caption{Normalised histograms of orbital inclinations and eccentricities for light and massive particles. Top panels show the cosines of orbital inclinations for 1X, 10X and 30X models. Bottom panels show eccentricities for the same models. Solid and dashed lines indicate massive ($m \ge 10\msun$) and light ($m < 10\msun$) particles respectively. The histograms correspond to the same time snapshots as in Fig.~\ref{fig:i-a-od}. The shaded histogram in the bottom right panel shows the initial eccentricity distribution for all of these models. } 
	\label{fig:hists-od}
\end{figure*}

\begin{figure*}
	\begin{center}
		\includegraphics[width=\linewidth]{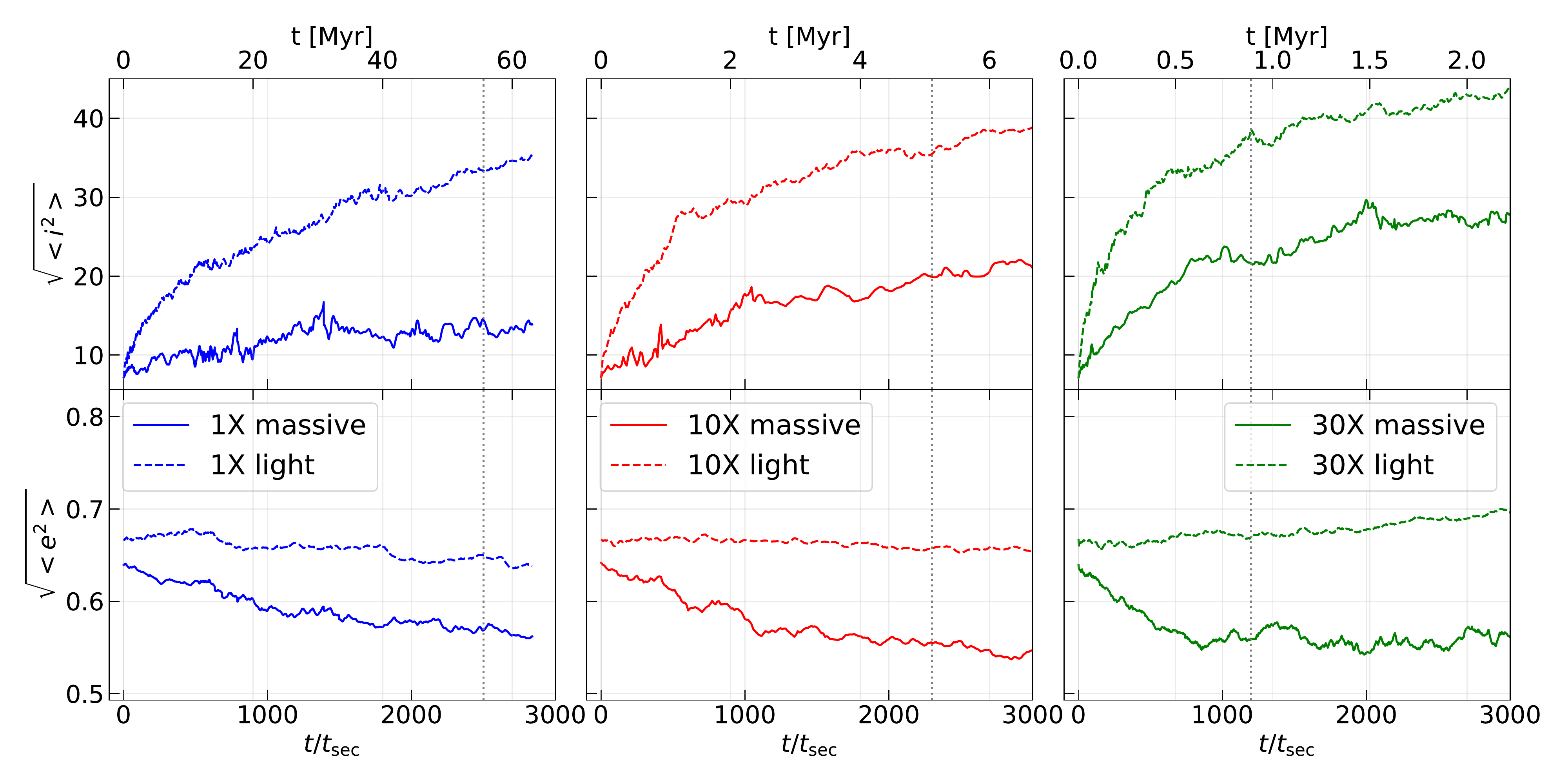} \\ 
		\par\end{center}
	\caption{Evolution of root-mean-square inclination angles and eccentricities for the same thermal isolated disc models as in Fig.~\ref{fig:i-a-od} and Fig.~\ref{fig:hists-od} a function of secular time (defined in Eq.~\ref{eq:tsec}). Solid and dashed lines indicate massive and light particles respectively. Dotted vertical lines show the time at corresponding to the snapshots of Fig.~\ref{fig:i-a-od} and  Fig.~\ref{fig:hists-od}. } 
	\label{fig:rms-od}
\end{figure*}

\begin{figure}
	\begin{center}
		\includegraphics[width=\columnwidth]{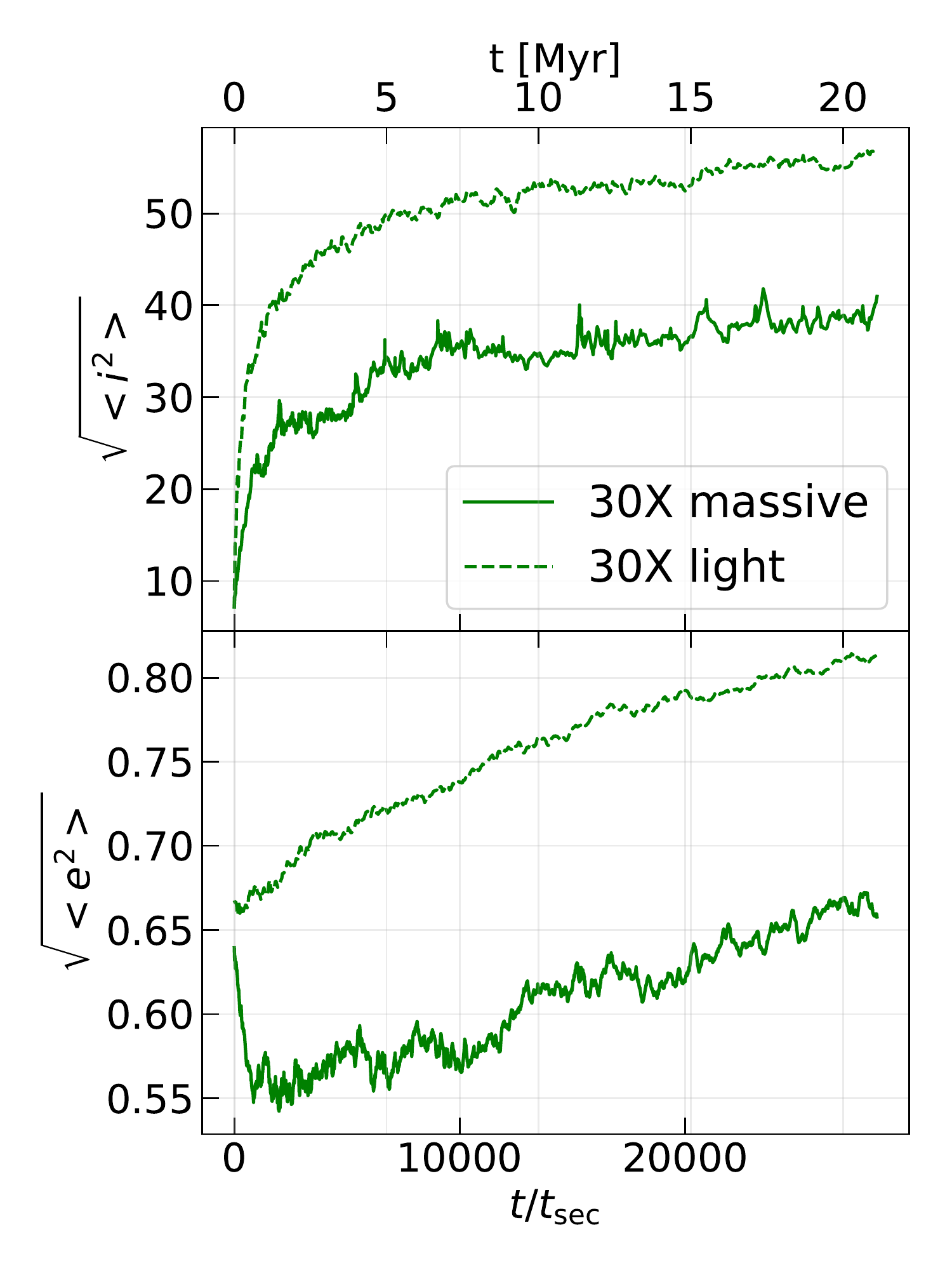} \\ 
		\par\end{center}
	\caption{Long-term evolution of the root-mean-square inclination angles and eccentricities for the 30X model. Top lines in each panel show light particles.} 
	\label{fig:x30_rms-od}
\end{figure}

\begin{figure}
	\begin{center}
		\includegraphics[width=\columnwidth]{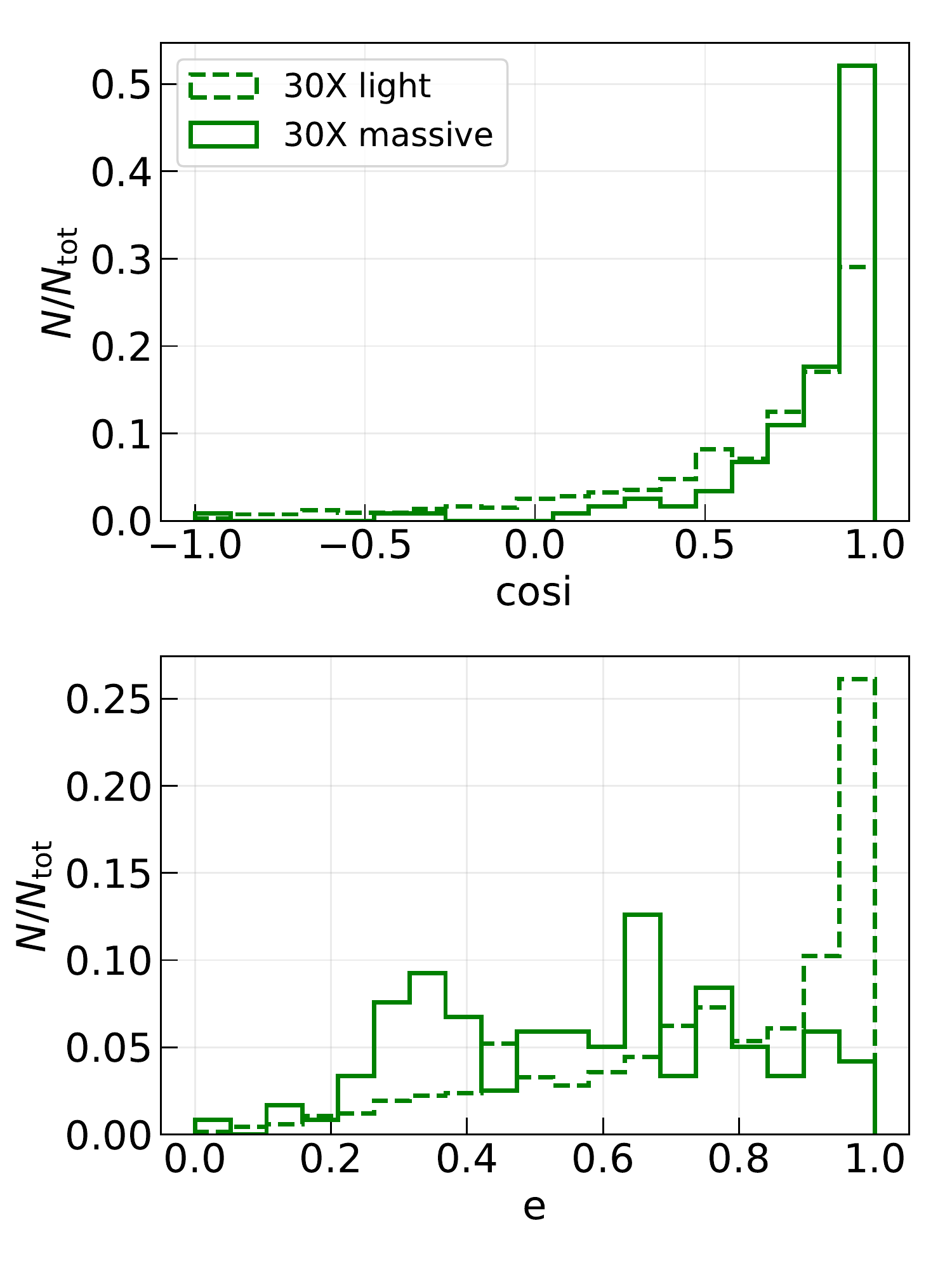} \\ 
		\par\end{center}
	\caption{Normalised histograms of the cosines of orbital inclinations and eccentricities for light and massive particles. For the 30X models at the moment of 28000 $t_\mathrm{sec}$. } 
	\label{fig:x30_hists-od}
\end{figure}

\begin{figure*}
	\begin{center}
		\includegraphics[width=\linewidth]{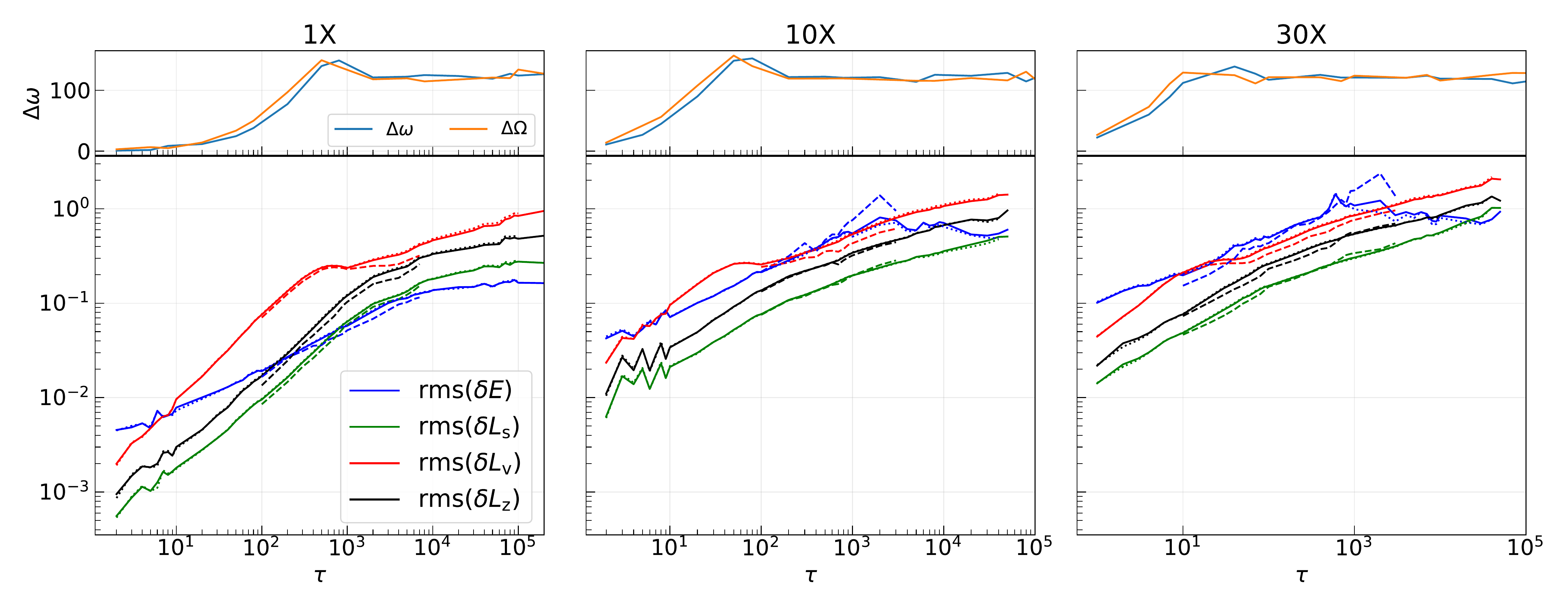} \\ 
		\par\end{center}
	\caption{The RMS change in the Keplerian energy (blue) and angular momentum (green, red, black) as a function of time for the $thermal$ isolated disc models. The angular momentum vector magnitude (green), angular momentum vector direction (red) and $Z$-component of the angular momentum vector (black) are shown as defined in Eqs.~(\ref{eq:dLz}). Solid, dashed, and dotted lines show all stars, massive stars ($m\geq 10\msun$), and light stars ($m\leq 10\msun$), respectively. The $Y$-axis shows the quantities in the legend, the $X$-axis shows the dimensionless time normalised to the orbital period (Eq.~\ref{eq:tau}). The top panels show the average change in the argument of periapsis ($\Delta\omega$) and the longitude of the ascending node ($\Delta\Omega$); the coherence time of SRR and VRR are related to the apsidal and nodal precession periods.} 
	\label{fig:rr-od}
\end{figure*}

The dynamical relaxation processes are expected to change the distribution of orbital inclination angles by warping, twisting, and affecting the thickness of the disc.
The left panel of Figure \ref{fig:i-a-od} shows the $10\%$, $50\%$, and $90\%$ cumulative distribution levels of orbital inclination angles as a function of semi-major axis. The innermost stars tend to have higher orbital inclinations, which is explained by the shorter relaxation time-scales at smaller distances from the SMBH (see Sec.~\ref{sec:time-scales}). The right panel of the figure shows the average inclination angle as a function of mass indicating that the high-mass stars (black holes) have systematically lower inclinations forming a thin disc. This effect develops in all models with an isolated stellar disc. The time instances corresponding to the 1X, 10X and 30X models in Fig.~\ref{fig:i-a-od} are chosen to have the same average inclination angle for the light stars ($m \leq 10 \msun$)\footnote{This choice is somewhat arbitrary, but as we see from the right panel of Fig.~\ref{fig:i-a-od} the value $m = 10 \msun$ is in the mass-gap produced by the stellar evolution and all objects with higher masses in the simulation are stellar mass black holes.}, implying that the curves in the right panel of Fig.~\ref{fig:i-a-od} overlap for light stars by construction. The inclination versus semi-major axis shows very similar trends in the left panel of the figure implying that all models are at the same level of relaxation. 

In the top panels of Fig.~\ref{fig:hists-od}, we compare the distribution of cosines of the orbital inclinations for massive ($m \geq 10 \msun$) and light ($m\leq 10\msun$) stellar objects. Each panel corresponds to the model with different mass factors (1X, 10X and 30X) at the same time as in Fig.~\ref{fig:i-a-od}. While the distribution of low mass stars are identical by construction, the 1X model clearly shows the strongest effect in vertical mass segregation compared to 10X and 30X models. Bottom panels of the same figure demonstrate that the isolated stellar discs also feature mass segregation in the eccentricity distribution as seen from the normalised distribution of orbital eccentricities for light and massive stars. But in this case the higher-mass models show stronger mass segregation than the 1X model. 

To examine the time dependence of mass segregation in inclination and eccentricity and its dependence on the 1X, 10X, 30X models, we track the time evolution of the root-mean-square (rms) inclination angles and eccentricities for all the models as a function of secular time. Fig.~\ref{fig:rms-od} confirms the expectation that vertical mass segregation is strongest for the 1X models and the weakest for the 30X models while mass dependence in eccentricities is the opposite. 

To explore the long-term evolution of isolated stellar discs, we focus on the 30X model which is numerically the least expensive. Fig.~\ref{fig:x30_rms-od} shows that massive and light stars develop a different rms inclination and eccentricity during the first stages of the evolution and continue with the same pace after a few thousand secular times. As a result, mass segregation effects are expected to be present in such systems (see Fig.~\ref{fig:x30_hists-od}). 

Vertical mass segregation in galactic nuclei may be caused by vector resonant relaxation as shown first by \citet{Szolgyen2018} and later confirmed by other studies \citep{Fouvry2020,Magnan2021,Mathe2022}. On the other hand, angular momentum conservation during pairwise interactions implies that two-body relaxation may also cause vertical mass segregation in the long-run especially in highly anisotropic systems \citep{Ernst+2007,Tiongco2021}. The mass segregation in eccentricities may be caused by both scalar resonant relaxation \citep{Fouvry2018,Gruzinov2020} and two-body relaxation. As shown by \citet{Alexander2007}, the rms eccentricity of a stellar disc is related to its velocity dispersion as:
\begin {equation}
e_{\rm{rms}} = \sqrt{2}\frac{\sigma}{v_\mathrm{K}},
\end {equation}
where $v_\mathrm{K}$ is the Keplerian orbital speed. Following this logic, \citet{Mikhaloff2017} showed that the evolution of rms eccentricities is different for light and heavy stars as a result of two-body interactions.

To explore which relaxation process drives anisotropic mass segregation predominantly in our models of isolated stellar discs with no spherical component, we perform the correlation curve analysis \citep{Rauch1996, Eilon2009, Kocsis2015}. We measure changes in energies and angular momenta for each particle to compute the rate of diffusion in energy -- angular momentum for the whole system (see Appendix~\ref{app:RR} for details). Fig.~\ref{fig:rr-od} shows the rms change in Keplerian energy (to track two-body relaxation), angular momentum magnitude (to track SRR), angular momentum vector direction (to track VRR) and the $Z$-component of the angular momentum vector (VRR in vertical direction) relative to the initial state.
Clearly, VRR strongly dominates in the 1X models: the relative change in angular momentum vector direction occurs faster than the change in other quantities (the red line is always above). However, due to the strong nodal precession, the change is predominantly along the azimuthal component of the angular momentum vector, while the orbital inclination is nearly constant. The mixing of orbital inclination angles is represented by the change in the $Z$-component of the angular momentum vectors (shown as a black line in Fig.~\ref{fig:rr-od}). This is suppressed initially compared to the change in the energy, but becomes more prominent after $10^2$ orbital periods. For 10X and 30X models two-body relaxation is the most efficient relaxation process, at least during the first $10^3$ periods. Fig.~\ref{fig:rr-od} also shows the comparison of the efficiency of the relaxation processes for massive (dashed lines of the same colour) and light (dotted lines of the same colour) stars. As light stars represent the majority of the system, they are almost indistinguishable from the overall cluster properties. On the other hand, the difference between the change in energy and angular momentum for massive stars indicates that the diffusion in energy and angular momentum for the massive stars is less efficient. Since relaxation is driven by VRR in the 1X model, this explains the strongest vertical mass segregation compared to 10X and 30X models discussed above (see Fig.~\ref{fig:rms-od}). On the other hand, energy and angular momentum changes in the 10X and 30X models are mostly driven by two-body relaxation. This explains the more prominent mass segregation effect in the eccentricities in 10X and 30X models compared to the 1X model. Since the contribution from SRR is the least significant for the studied models (especially the 10X and 30X models, see green lines in Fig.~\ref{fig:rr-od}), we conclude that the anisotropic mass segregation effects in $thermal$ isolated discs are caused by both VRR and two-body relaxation.

\section{Interaction of a nuclear stellar disc with a spherical cusp of stars}
\label{sec:disk-sphere-evol}

\begin{figure*}
	\begin{center}
		\includegraphics[width=\linewidth]{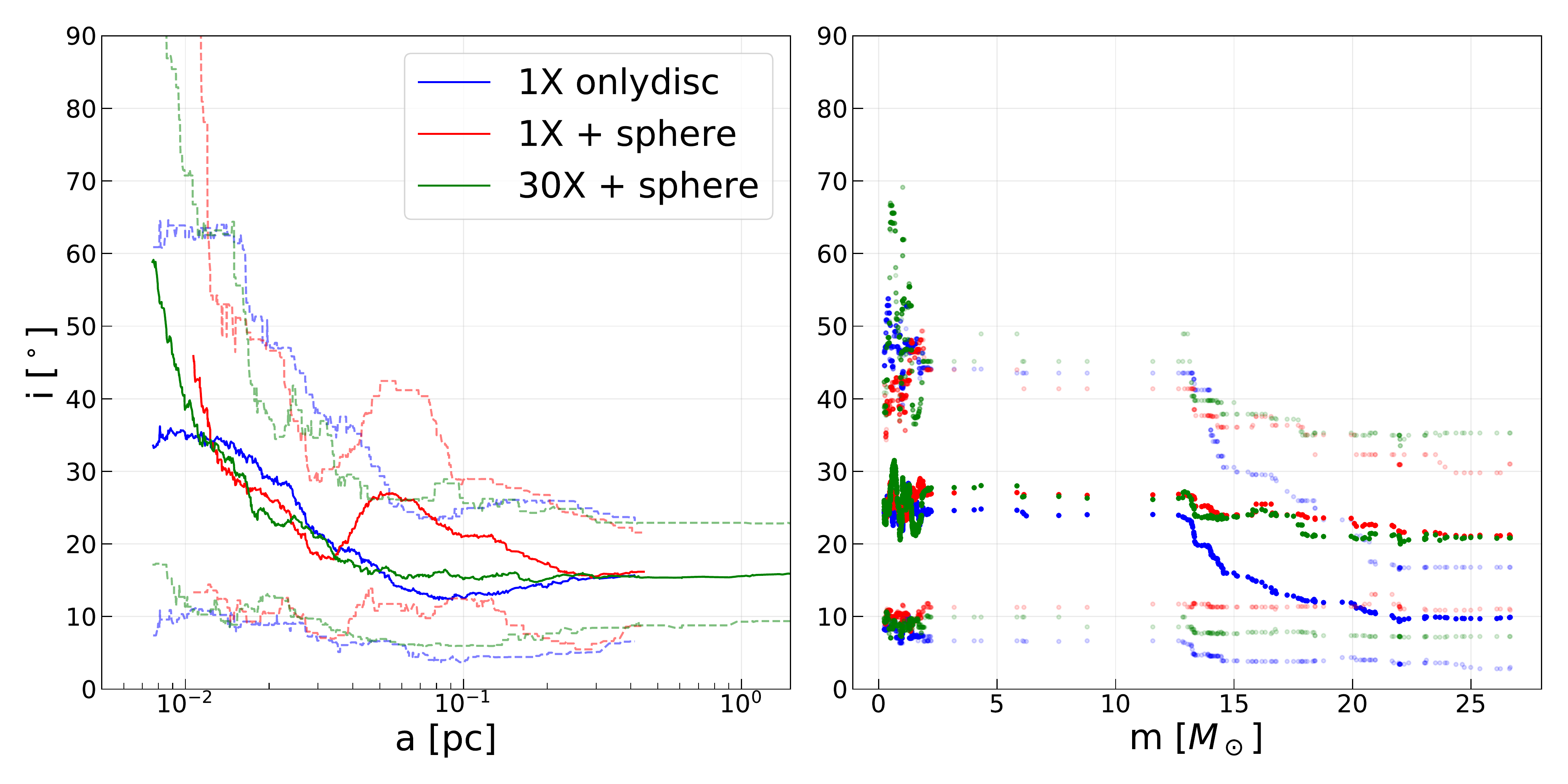} \\ 
		\par\end{center}
	\caption{Similar to Fig.~\ref{fig:i-a-od} but also showing models with and without a spherical component. Left panel: moving average of inclination angles in semi-major axes. Colours show models with different mass factors (see legend). Faded lines (points) of the same colour show 90 and 10\% quantiles. The time snapshots shown are chosen such that the average orbital inclination are identical for the low mass stars (this corresponds to 40, 2.6 and 0.033 Myr for 1X only disc model, the 1X disc+sphere and 30X disc+sphere models respectively; see dotted lines in Fig.~\ref{fig:rms-sphere}). Blue lines in both panels indicate the isolated disc with the \textit{stardisc} initial conditions and the power-law density slope $\gamma=3.3$. Red and green lines show the same disc embedded in the spherical stellar cusp with the power-law slope of $\gamma=1.75$ for 1X and 30X models, respectively (see Sec.~\ref{sec:simulations} for detailed description of the models). }
	\label{fig:i-a-sphere}
\end{figure*}

\begin{figure*}
	\begin{center}
		\includegraphics[width=\linewidth]{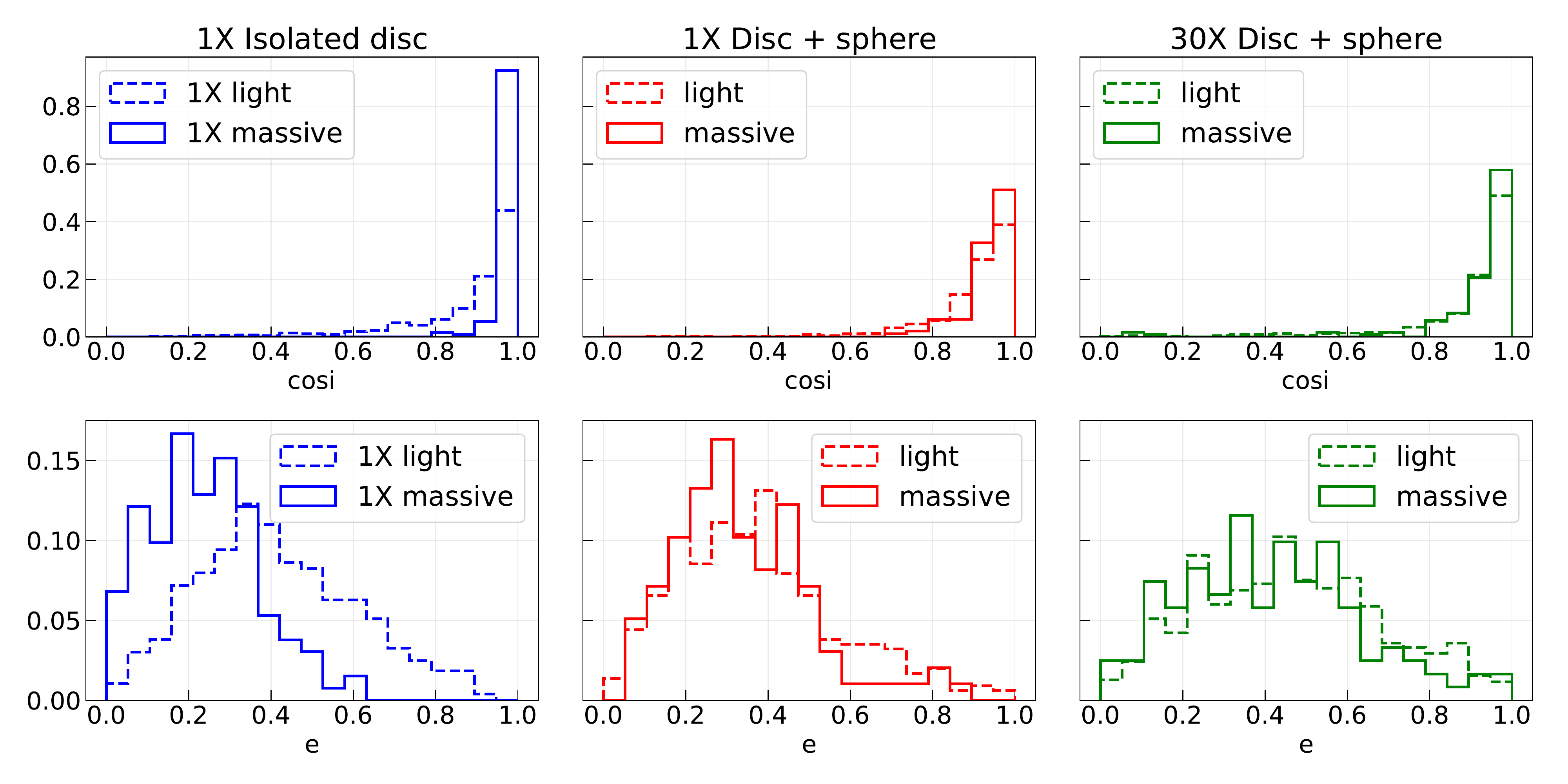} \\ 
		\par\end{center}
	\caption{Normalised histograms of orbital inclinations and eccentricities for light ($m < 10\msun$) and massive ($m\ge 10\msun$) particles to compare the isolated disc model with models that feature a spherical stellar cusp for the same models as in Fig.~\ref{fig:i-a-sphere}. } 
	\label{fig:hists-sphere}
\end{figure*}

We analyse the shape and thickness of the stellar disc using the quadrupole moment matrix (see e.g. \citealt{Roupas2017}, \citealt{Szolgyen2021}) defined as follows: 

\begin{equation}\label{eq:Q}
\boldsymbol{\mathrm{Q}}_{\alpha\beta}=\dfrac{\sum_{i=1}^N L_{i\alpha}L_{i\beta}}{\sum_{i=1}^N{|\boldsymbol{L}_{i}|^2}},
\end{equation}
where $\boldsymbol{L}_{i}$ is the angular momentum vector of the $i$-th star, $\alpha$ and $\beta$ are the corresponding Cartesian components.

The largest eigenvalue of the matrix corresponds to the shape of the disc while the corresponding principal eigenvector describes the orientation of the system 
\begin{equation}\label{eq:Eigenvectors}
\boldsymbol{\mathrm{Q}}_{\alpha\beta}\boldsymbol{\nu}=\lambda\boldsymbol{\nu}.
\end{equation}
In this normalisation, the trace of the matrix satisfies $\mathrm{Tr} \boldsymbol{\mathrm{Q}} = 1$ meaning that equal eigenvalues $\lambda_1=\lambda_2=\lambda_3 = \frac13$ represent a sphere with zero angular momentum, and a razor-thin disc has $(\lambda_1,\lambda_2,\lambda_3) = (1,0,0)$. Thus, the largest eigenvalue which takes the values $1/3 \le \lambda \le 1$ quantifies the thickness of the stellar disc.

In this section, the inclination angles of the disc stars refer to the mean inclinations with respect to the principal eigenvector of the disc. This way, the orbital inclinations are always computed relative to the instantaneous mid-plane of the disc in angular momentum space even if the disc as a whole is tilted with respect to its initial position.

\begin{figure*}
	\begin{center}
		\includegraphics[width=\linewidth]{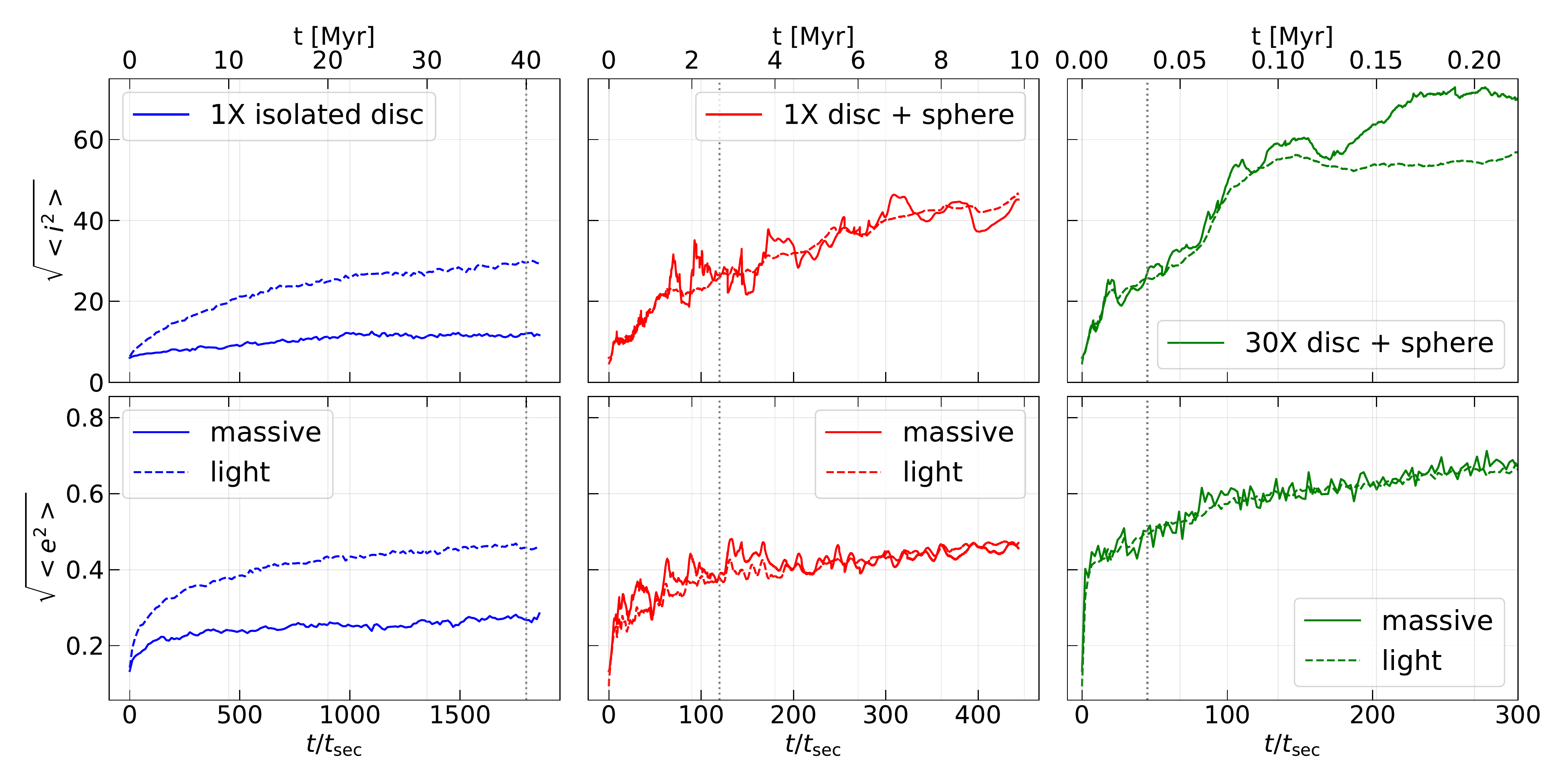} \\ 
		\par\end{center}
	\caption{Time-evolution of RMS eccentricities and inclination angles for 1X isolated disc, 1X disc + sphere, and 30X disc + sphere models as a function of secular time for massive (solid lines) and light (dashed lines) particles for the same models as in Fig.~\ref{fig:i-a-sphere} and  \ref{fig:hists-sphere}. } 
	\label{fig:rms-sphere}
\end{figure*}

\subsection{Secular evolution of the embedded nuclear stellar discs}
\label{subsec:secular}

We follow the same steps as in Sec.~\ref{sec:disk-evol} to study the dynamics of discs embedded in a spherical cusp of stars on secular time-scales. But in this subsection we use the \textit{stardisc} models with the power-law slope $\gamma=3.3$ and compare the isolated disc, the disc embedded in a sphere and the 30X model of the same disc embedded in a sphere. 
The semi-major axes -- inclination dependence (left panel of Fig.~\ref{fig:i-a-sphere}) is qualitatively similar for all 3 models, but the models with a spherical component extend to higher inclinations in the innermost part. 

The right panel of Fig.~\ref{fig:i-a-sphere} shows a striking difference in the average inclinations as a function of mass for high-mass stars: while the isolated disc model shows lower inclination angles with increasing mass, there is almost no correlation between stellar mass and orbital inclinations for models with an isotropic spherical component. Similar to Fig.~\ref{fig:i-a-od}, the time instances shown in Fig.~\ref{fig:i-a-sphere} have the same average inclination angle for the low-mass stars ($m\leq 10\msun$) by construction. While the vertical mass segregation effect vanishes, the dependence of the inclination on the semimajor axis is more prominent. The latter effect develops faster and extends to higher inclinations, and some stars even flip to counter-rotating orbits ($i > 90^\circ$).

\begin{figure*}
	\begin{center}
		\includegraphics[width=\linewidth]{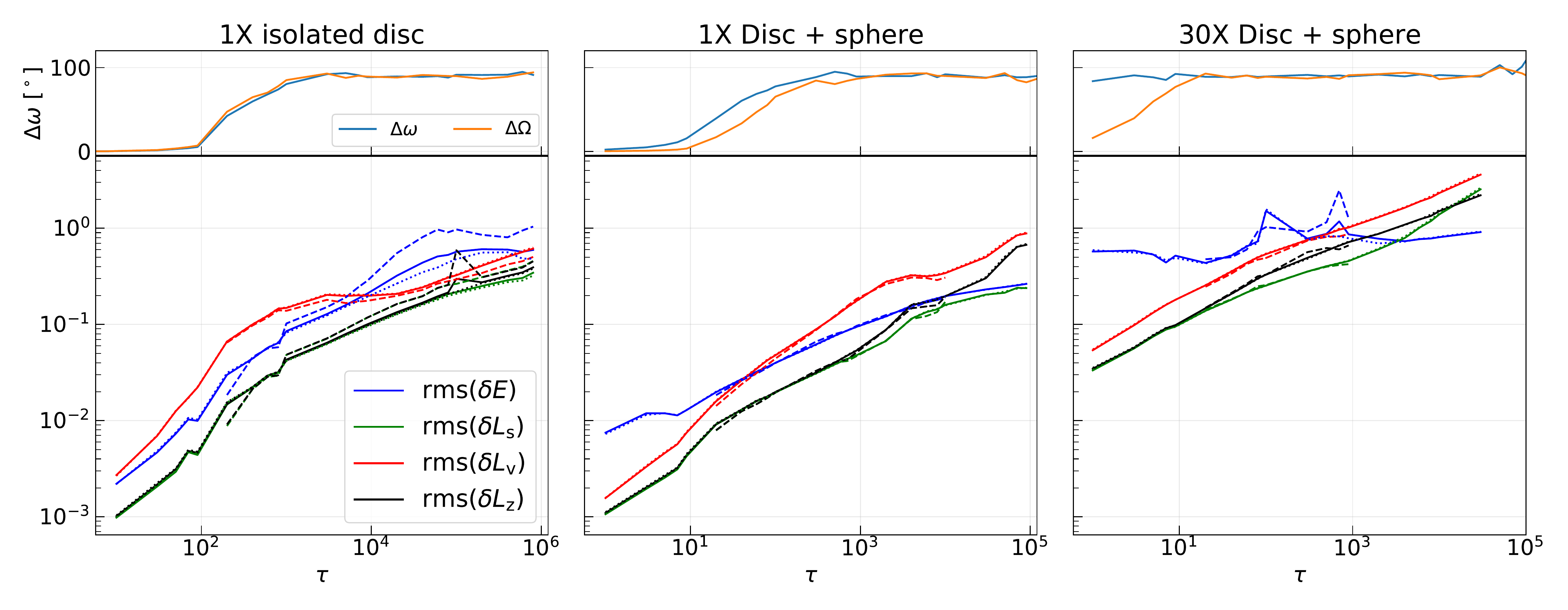} \\ 
		\par\end{center}
	\caption{Similar to Fig.~\ref{fig:rr-od} but also showing models with and without a spherical component. Left panel shows the \textit{stardisc} isolated disc model with $\gamma=3.3$, the middle and right panels show 1X and 30X models of the same disc embedded in an isotropic stellar component. The values are measured only for the stars that initially belong to the disc. } 
	\label{fig:rr}
\end{figure*}

The top panels of Fig.~\ref{fig:hists-sphere} show the normalised distributions of light and massive particles respectively at the time-snapshots of Fig.~\ref{fig:i-a-sphere} demonstrating that the relative difference in high mass stars at low orbital inclinations is not more than 10\% than that for the light stars for the models with an isotropic spherical component. 
The bottom panels of Fig.~\ref{fig:hists-sphere} provide a comparison for the distribution of orbital eccentricities for massive and light stars between the three reference models. As in the case of orbital inclinations, mass segregation in eccentricities vanishes when the same stellar disc interacts with a spherical nearly isotropic distribution of stars. The time-evolution of the rms inclinations and eccentricities (Fig.~\ref{fig:rms-sphere}) shows that massive and light stars relax at the same rate. 

To understand which relaxation process dominates in these systems we examine the correlation curves for the relative changes in energies and angular momenta as for the case of an isolated disc discussed above (see Appendix~\ref{app:RR} for details). Fig.~\ref{fig:rr} shows that the isolated disc case (the left panel) is initially dominated by two-body relaxation. This is unsurprising as the initial condition for \textit{stardisc} models feature low eccentricities and low inclinations implying faster two-body relaxation initially (see Eq.~\ref{eq:t-2b-disk} and \citealt{Subr2014}). As the isolated disc system is highly anisotropic, we see that the internal dynamics leads to anisotropic mass segregation which in this case is driven by two-body relaxation (cf. Fig.~\ref{fig:rr-od} for the \textit{thermal} model showing less prominent energy diffusion). The middle panel of Fig.~\ref{fig:rr} shows the energy and angular momentum correlation curves for the disc embedded in a dominant isotropic spherical component. Here we do not observe any differences between the curves of the massive and light stars. Contrary to the case of the isolated disc, the 1X model with a spherical component shows that two-body relaxation dominates only in the initial phase of evolution (first $10^3$ periods) after which VRR takes over. After $10^4$ periods VRR fully dominates the evolution. We note that only the innermost particles contribute to the curves after $\tau = 10^4$ showing that the inclination -- semi-major axis anticorrelation presented in the left panel of Fig.~\ref{fig:i-a-sphere} is mostly driven by VRR. 

Applying this to the Milky Way galactic centre, $10^4$ orbital periods corresponds to less than 5 Myr for the stars with semi-major axes $a  < 0.05$ pc meaning that the $S$-stars are subject to an efficient VRR. The upper panels of the Fig.~\ref{fig:rr} show the average change in the argument of periapsis ($\omega$) and longitude of the ascending node ($\Omega$). As also expected from theory \citep{Rauch1996}, the figure shows that the coherent phase of SRR occurs on the apsidal precession time-scale. Further, as we suggested in Sec.~\ref{sec:time-scales}, for stellar discs embedded in a spherical component, the coherent phase of VRR takes place on the nodal precession time-scales. We refer to Appendix~\ref{app:RR} for a detailed analysis of the VRR efficiency. Contrary to the 1X model, the 30X models are dominated by two-body relaxation which takes place in $10^3$ orbital periods. Since the 30X models in our simulations are equivalent to dwarf galaxies with central black holes with masses of order $M_\mathrm{bh}\simeq10^5\msun$, we conclude that these systems are dominated by two-body dynamics. We explore such systems further in Sec.~\ref{subsec:long}.  

\subsection{Comparison to previous models}

To understand why stellar discs with an isotropic spherical component do not show a vertical mass segregation, while previous studies with nearly spherical initial conditions did show this effect \citep{Szolgyen2018, Magnan2021, Mathe2022}, we examine the dimensionless VRR energy and angular momentum in our models which determine the VRR equilibria as shown in \citet{Mathe2022}:
 \begin{align}\label{eq:Etotnorm}
    E_{\rm tot} = -\dfrac{\sum_{ij}\sum_{\ell=2}^{\ell_{\max}} \mathcal{J}_{ij\ell}P_{\ell}\left(\hat{\bm{L}}_i\cdot\hat{\bm{L}}_j\right)}{\sum_{ij}\sum_{\ell=2}^{\ell_{\max}} \mathcal{J}_{ij\ell}}\,,\quad
    L_{\rm tot} = \dfrac{\left|\sum_{i=1}^{N} \bm{L}_i\right|}{\sum_{i=1}^{N}|\bm{L}_i|}\,.
\end{align}
Here $\hat{\bm{L}}_i, \hat{\bm{L}}_j$ are units vectors in angular momentum direction for the $i^\mathrm{th}$ and $j^\mathrm{th}$ particles, $\ell$ is the multipole index, $\mathcal{J}_{ij\ell}$ are pairwise coupling coefficients that depend on eccentricities and semimajor axes and $P_{\ell}(x)$ are Legendre polynomials. Here, $(E_{\rm tot}, L_{\rm tot})=(0,0)$ represents an isotropic distribution, while $(-1,(1+\kappa)^{-1})$ corresponds to a razor thin disc where a $\kappa$ fraction of stars orbit in one sense and $1-\kappa$ in the other. We refer to \citet{Mathe2022} and \citet{Kocsis2015} for details.  

In our simulations, $(E_{\rm tot}, L_{\rm tot})$ are of order $(-10^{-4},10^{-4})$ for the models with a spherical component which are clearly very nearly isotropic. In comparison, the most isotropic case presented in \citet{Mathe2022} had $(E_{\rm tot}, L_{\rm tot})=(-0.03,0.16)$ which is relatively more anisotropic. Moreover, the models with dominating disc (presented in Sec.~\ref{subsec:sphere}) which are highly anisotropic do show vertical mass segregation. Thus, we conclude that the absence of anisotropic mass segregation in our models with an isotropic spherical component does not contradict previous studies of VRR, but it indicates that the final state of vertical mass segregation depends strongly on the deviation from isotropy. 
Note that two-body relaxation may also drive anisotropic mass segregation on the longer two-body relaxation timescale, but similarly to VRR, only in cases with an initial anisotropy in angular momentum vector space \citep[see][for related studies in globular clusters]{Tiongco2021,Livernois2022}. 
Thus, we conclude that vertical mass segregation is absent in our models with a disc+spherical component due to a very low net initial anisotropy; and our models of isolated discs (Sec.~\ref{sec:disk-evol})  exhibit anisotropic mass segregation as found in \citet{Mathe2022}.

\subsection{Long-term evolution of embedded nuclear stellar discs}
\label{subsec:long}

The long-term evolution is shaped by two-body interactions which may lead to the exchange of energy and angular momentum between the particles in the disc and sphere. In this subsection we focus on the 30X models which are dominated by two-body interactions and are numerically relatively inexpensive to study the long-term evolution on the two-body relaxation timescale. As we have shown in previous sections, two-body relaxation is relatively subdominant in the Galactic centre and the 30X models are not appropriate in that case. The 30X models represent one-to-one simulations of nuclear star clusters in dwarf galaxies with SMBHs of mass $M_\mathrm{bh}=4\times 10^6\msun/30 = 1.3\times 10^5\msun$
(see \citealt{Nguyen2019} for examples of galaxies hosting nuclear star clusters with massive black holes below $10^6\msun$). In the analysis below, we simulate the system with this SMBH mass and nuclear star clusters extending up to 1 pc. We estimate the two-body relaxation time using the half-mass relaxation time of the spherical component (Eq.~\ref{eq:t-2body}) using data from our simulations.

\begin{figure}
	\begin{center}
		\includegraphics[width=\columnwidth]{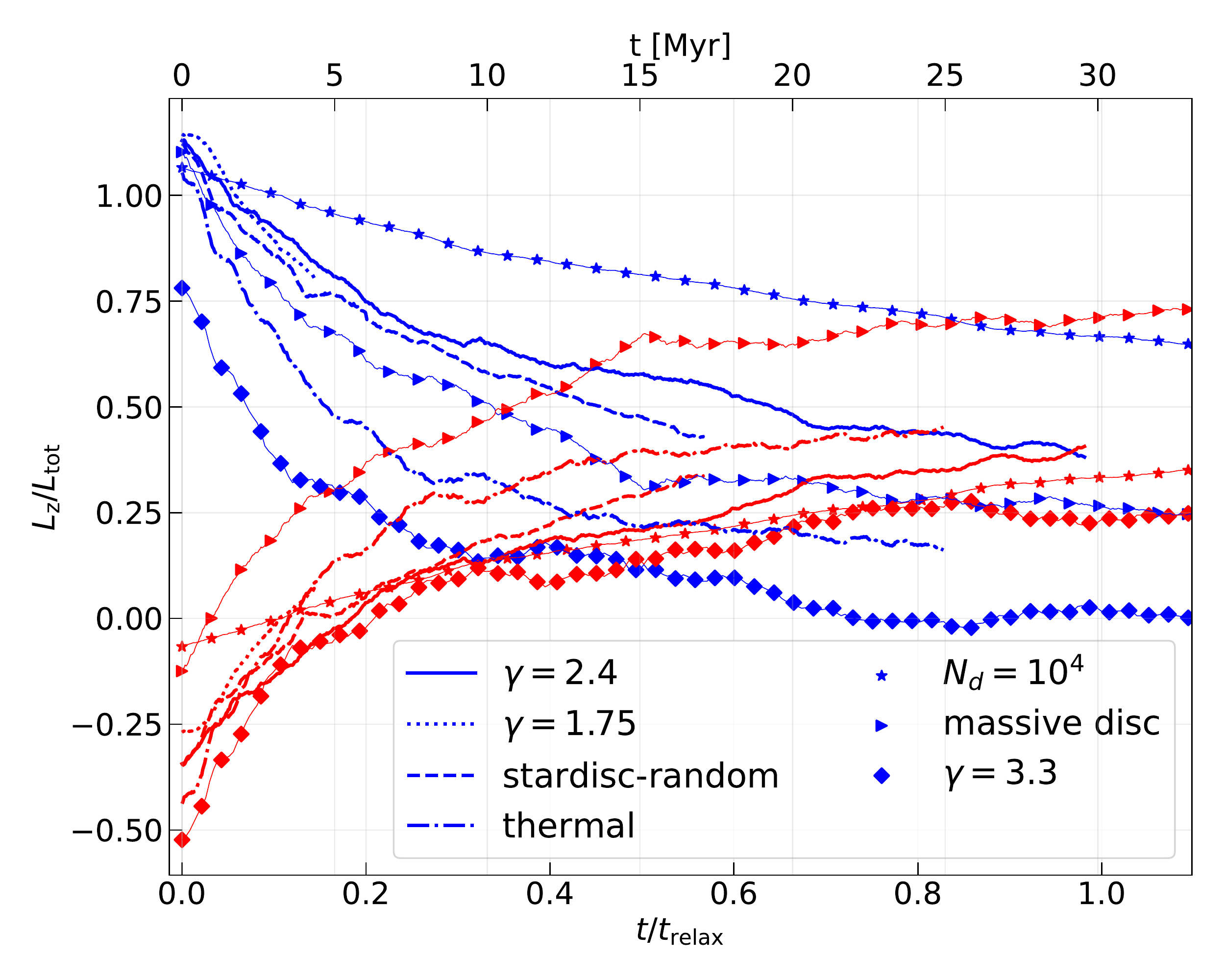} \\ 
		\par\end{center}
	\caption{$Z$-components of the angular momentum vectors of the disc (blue lines) and the sphere (red lines) normalised to the total angular momentum of the whole system. Bottom $X$-axis shows time in units of half-mass two-body relaxation time of a spherical component. Line styles correspond to different models of the stellar disc according to the legend. Shows only 30X models which are equivalent to a dwarf galaxy hosting a $1.3\times10^5$ black hole and a nuclear star cluster extended to 1 pc.}
	\label{fig:lz}
\end{figure}

Fig.~\ref{fig:lz} illustrates the exchange of the $z$-component of the angular momentum between the disc ($L_\mathrm{z,disc}$, blue curves) and the sphere ($L_\mathrm{z,sphere}$, red curves) showing their time-evolution normalised to the total angular momentum of the entire system ($L_\mathrm{tot}$) for the 30X models (see Table~\ref{tab:runs}) normalised to the total angular momentum of the entire system. The sphere has a nonzero initial $L_\mathrm{z,sphere}$ due to shot-noise-type stochastic deviation from isotropy, i.e. the initial value of $L_\mathrm{z,sphere}/L_{\rm z,disc}$ is drawn from a uniform distribution between $\pm \langle N m^2\rangle_{\rm sphere}^{1/2}/\langle N m\rangle_{\rm disc}= (N_{\rm sphere}^{1/2}/N_{\rm disc}) (\langle m^2\rangle_{\rm sphere}^{1/2}/\langle m\rangle_{\rm disc})$ where $m$ is the stellar mass. The disc tends to give away its angular momentum until it is completely mixed with the spherical component, i.e. when the net $L_\mathrm{z}$ per particle is equal for the two components, i.e.
\begin{equation}
\label{eq:Ld-mix}
       \frac{L_\mathrm{z,disc}}{L_\mathrm{tot}} \to \frac{\left<L_\mathrm{z,tot}\right>N_\mathrm{disc}}{L_\mathrm{tot}}\,,\quad
        \frac{L_\mathrm{z,sphere}}{L_\mathrm{tot}} \to \frac{\left<L_\mathrm{z,tot}\right>N_\mathrm{sphere}}{L_\mathrm{tot}}\,.
\end{equation}
Although, none of the simulations reached complete mixing, Fig.~\ref{fig:lz} demonstrates that $L_\mathrm{z,disc}$ approaches the equilibrium value of Eq.~\eqref{eq:Ld-mix} for all models. 

\begin{figure}
	\begin{center}
		\includegraphics[width=\columnwidth]{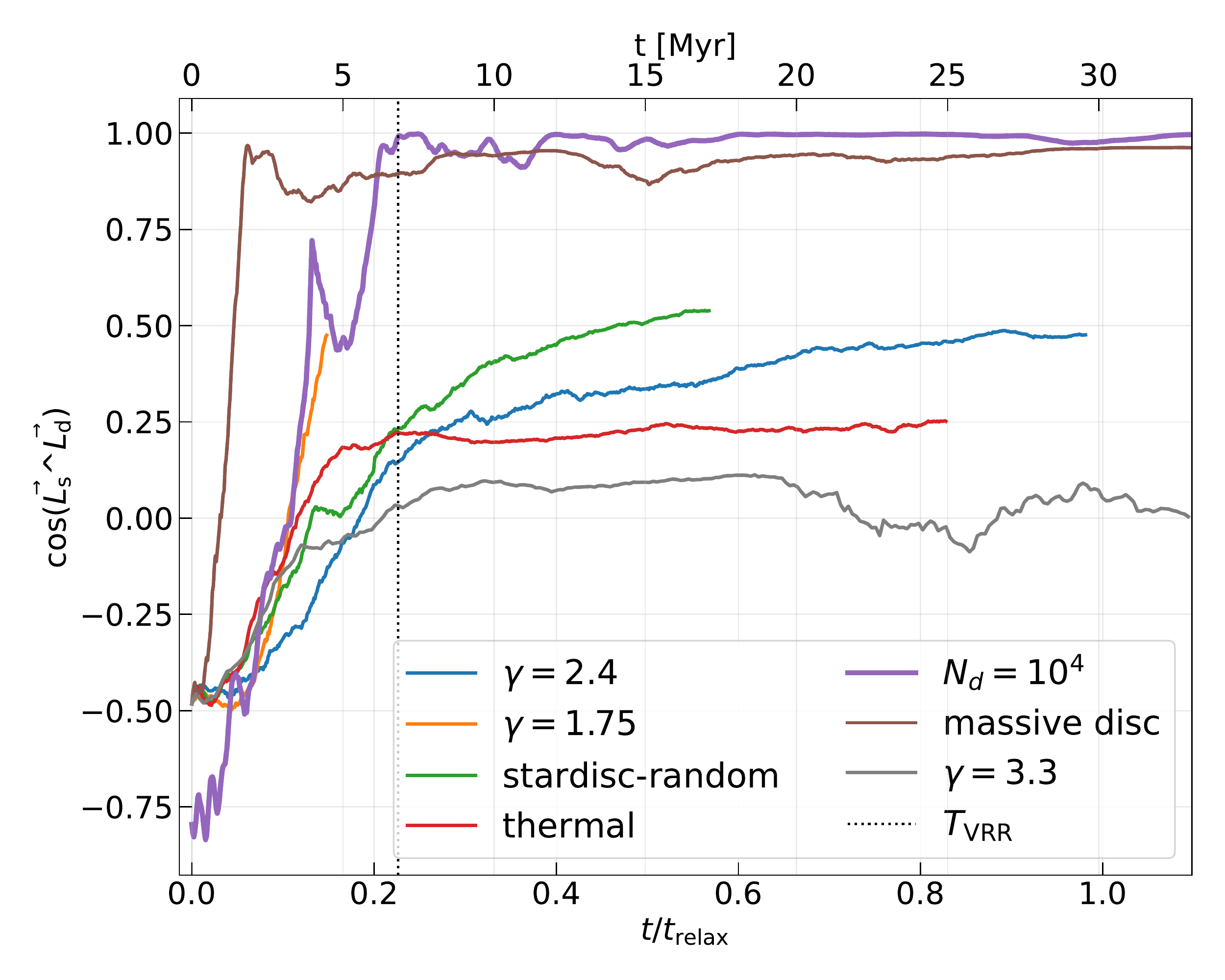} \\ 
		\par\end{center}
	\caption{Orientation of the total angular momentum vectors of the stellar disc and sphere, respectively, for the 30X models as in Fig.~\ref{fig:shape}, i.e. the cosine of the angle between the respective total angular momentum vectors as a function of time in units of half-mass two-body relaxation time of the spherical component. Line colours correspond to different disc models as indicated in the legend. The dotted vertical line represents the vector resonant relaxation time due to the spherical stellar cusp (Eq.~\ref{eq:t-vrr}). }
	\label{fig:orient}
\end{figure}

Fig.~\ref{fig:orient} illustrates the alignment of the respective total angular momentum vectors of the disc and spherical components in our simulations. Alignment occurs if the stellar disc is massive enough,  $N_{\rm disc}\langle m\rangle_{\rm disc}\gg N_{\rm sphere}^{1/2}\langle m^2\rangle^{1/2}_{\rm sphere}$, and if so, alignment takes place within the vector resonant relaxation time-scale shown by a vertical dotted line. For lower disc masses, $\bm{L}_{\rm disc}$ and $\bm{L}_{\rm sphere}$ end up in the same hemisphere (cosine of the mutual inclination angle is positive) even if they were counter-rotating initially as seen in Fig.~\ref{fig:lz} where both the disc and the sphere attain a net positive angular momentum.

\begin{figure}
	\begin{center}
		\includegraphics[width=\columnwidth]{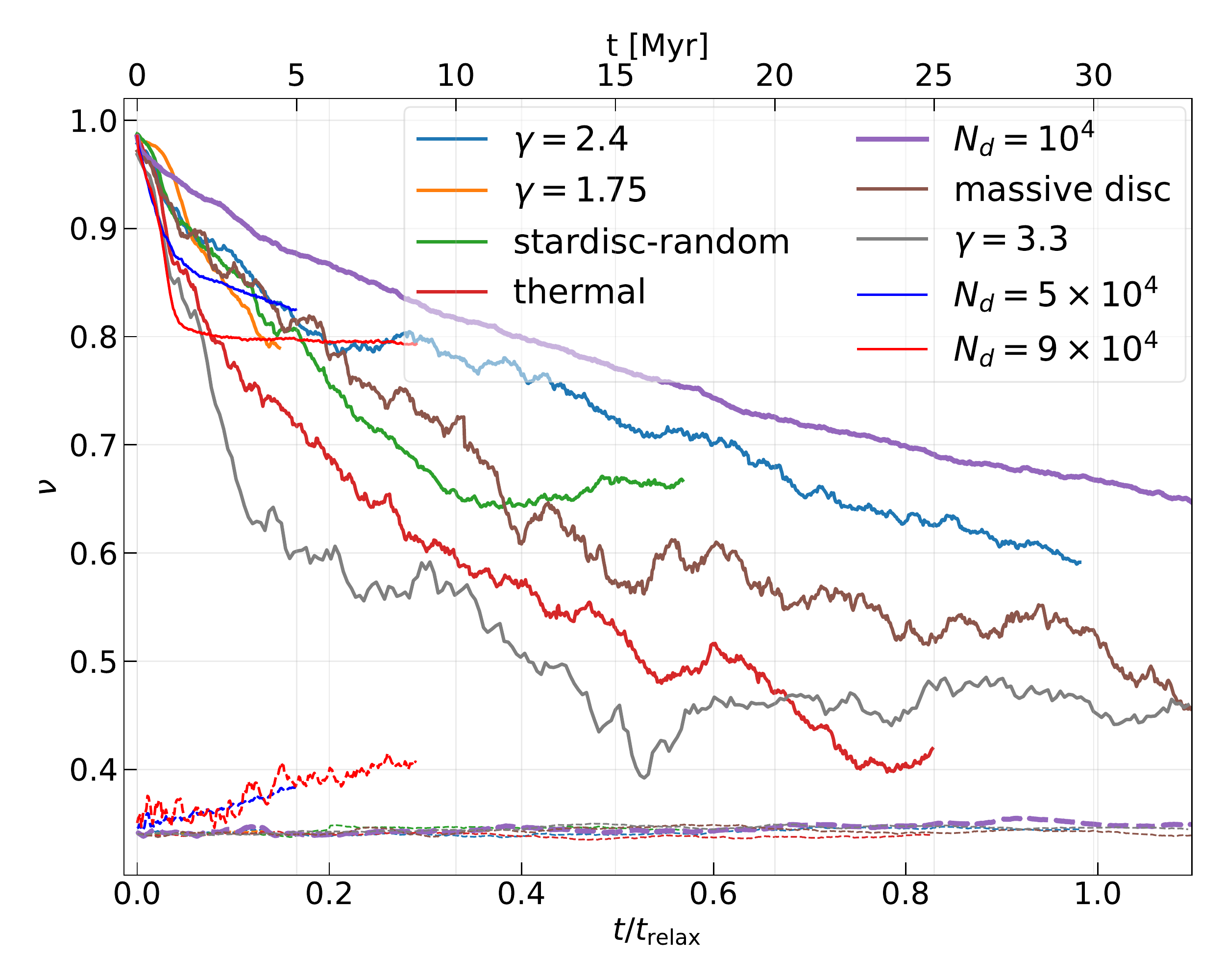} \\ 
		\par\end{center}
	\caption{Time-evolution of the thickness of the stellar disc (top solid lines) in units of half-mass two-body relaxation time of a spherical component quantified by the largest eigenvalue of the quadrupole moment matrix (see Eq.\ref{eq:Q}). $\nu=1$ describes the razor-thin disc while $\nu=1/3$ indicates the spherically symmetric distribution. The bottom dashed lines show the corresponding eigenvalue of the sphere. Line colours are the same as in Fig.~\ref{fig:orient} with additional blue and red lines corresponding to the disc-dominated models with number of stars in the disc $N_\mathrm{d}=5\times10^4$ and $N_\mathrm{d}=9\times10^4$. }
	\label{fig:shape}
\end{figure}

\begin{figure}
	\begin{center}
		\includegraphics[width=\columnwidth]{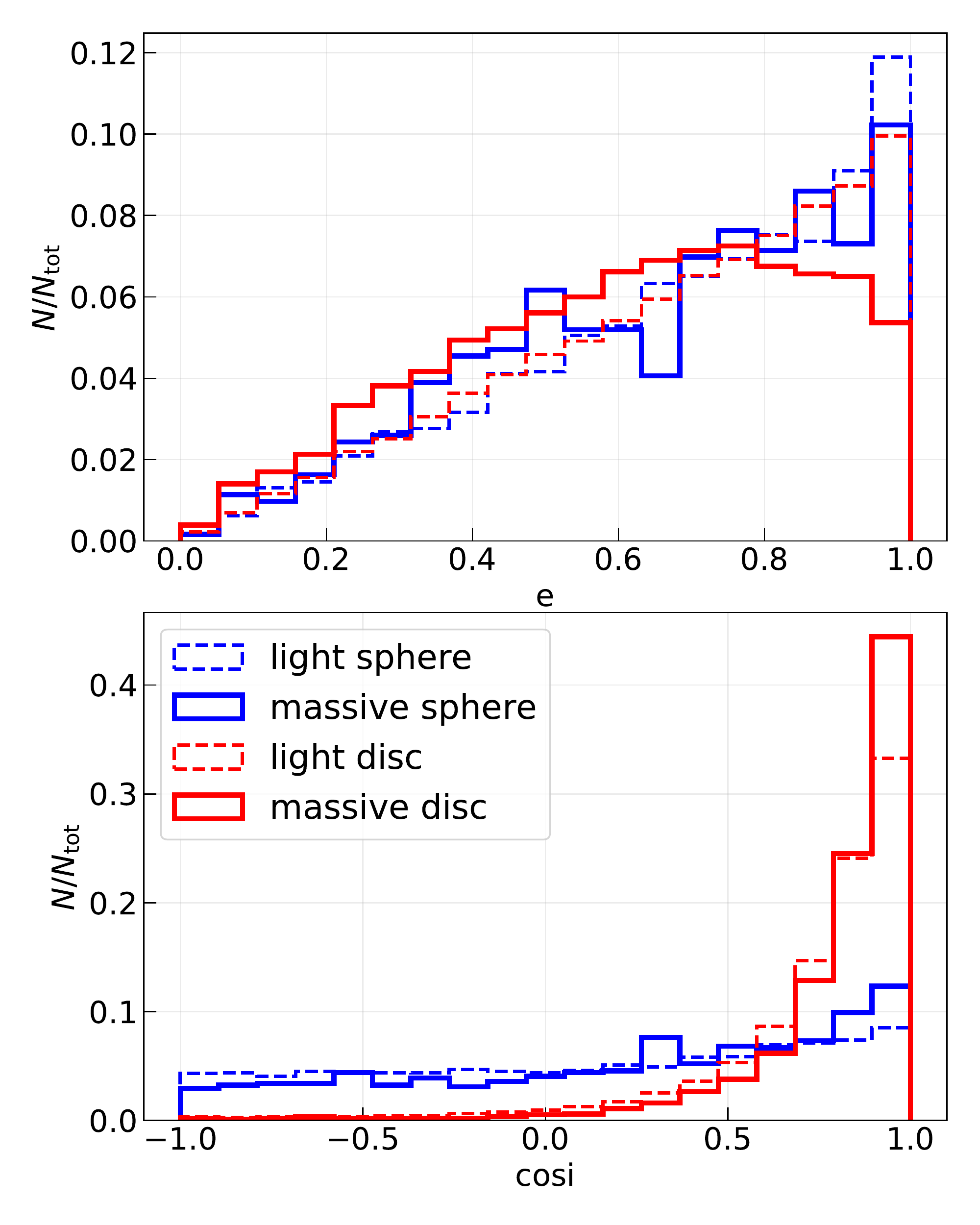} \\ 
		\par\end{center}
	\caption{Normalised histograms of eccentricities (top panel) and cosines of orbital inclinations (bottom panel) for the dominating disc model with $N_\mathrm{d}=9\times10^4$. Solid lines show the distribution of the massive stars, dashed lines correspond to the light stars. Blue lines show the stars in the sphere and red lines show the stars that were originally in the disc. 
	Shown at $t/t_{\rm relax}\simeq 0.3$.}
	\label{fig:dom_disk}
\end{figure}

Fig.~\ref{fig:shape} shows the evolution of the shape of the stellar disc quantified by the largest eigenvalue of the quadrupole moment matrix (defined in Eq.~\ref{eq:Q}) as a function of time. Generally, the angular momentum transfer from the stellar disc to a spherical component results in the thickening of the disc. Eventually, the disc appears to evolve towards a spherical shape. 

\subsection{Effect on the sphere}
\label{subsec:sphere}

As we have seen in the previous subsection, the stellar disc tends to evolve towards an isotropic distribution while interacting with the isotropic spherical star cluster. At the same time, as the spherical component absorbs the angular momentum of the disc, it preserves its original shape as long as it is much more massive than the disc (dashed lines in Fig.~\ref{fig:shape}). This is the case for relatively low-mass stellar discs (at most 15\% of the total stellar mass in our models), but in the case of the disc dominated models ($N_\mathrm{d}=0.5N_\mathrm{tot}$ and $N_\mathrm{d}=0.9N_\mathrm{tot}$) the angular momentum vector distribution flattens significantly for the initially isotropic sphere on the two-body relaxation time-scales. The upper limit for the degree of flattening attained by a spherical component may be determined from the total angular momentum budget of the whole system (see Eq.~\ref{eq:Ld-mix}). A similar conclusion was reached in \citet{Mastrobuono-Battisti2013,Mastrobuono-Battisti2016} for simulations of globular clusters which also flatten due to angular momentum transfer from a stellar disc, especially in case the disc mass exceeds $\sim$25\% of the total mass of the cluster.

Finally, Fig.~\ref{fig:dom_disk} demonstrates that when the disc is massive enough to cause flattening of a spherical component both the disc and spherical components feature vertical mass segregation. This is in line with expectations from VRR dynamics: the total energy -- total angular momentum pairs for the dominating disc models (evaluated using Eq.~\ref{eq:Etotnorm}) are ($E_{\rm tot}, L_{\rm tot}) = (-0.24, 0.46)$ and $(E_{\rm tot}, L_{\rm tot}) = (-0.65, 0.87)$ for the models with $N_\mathrm{d}=0.5\,N_\mathrm{tot}$ and $N_\mathrm{d}=0.9\,N_\mathrm{tot}$ implying a large amount of initial anisotropy. However, note that these 30X models are predominantly driven by two-body relaxation. Furthermore, these models also develop a mass segregation in eccentricity space (top panel in Fig.~\ref{fig:dom_disk}). These models show that two-body relaxation also plays an important role in driving anisotropic mass segregation.

\section{Application to the Galactic centre $S$-stars}
\label{sec:s-stars}

\begin{figure}
	\begin{center}
		\includegraphics[width=\columnwidth]{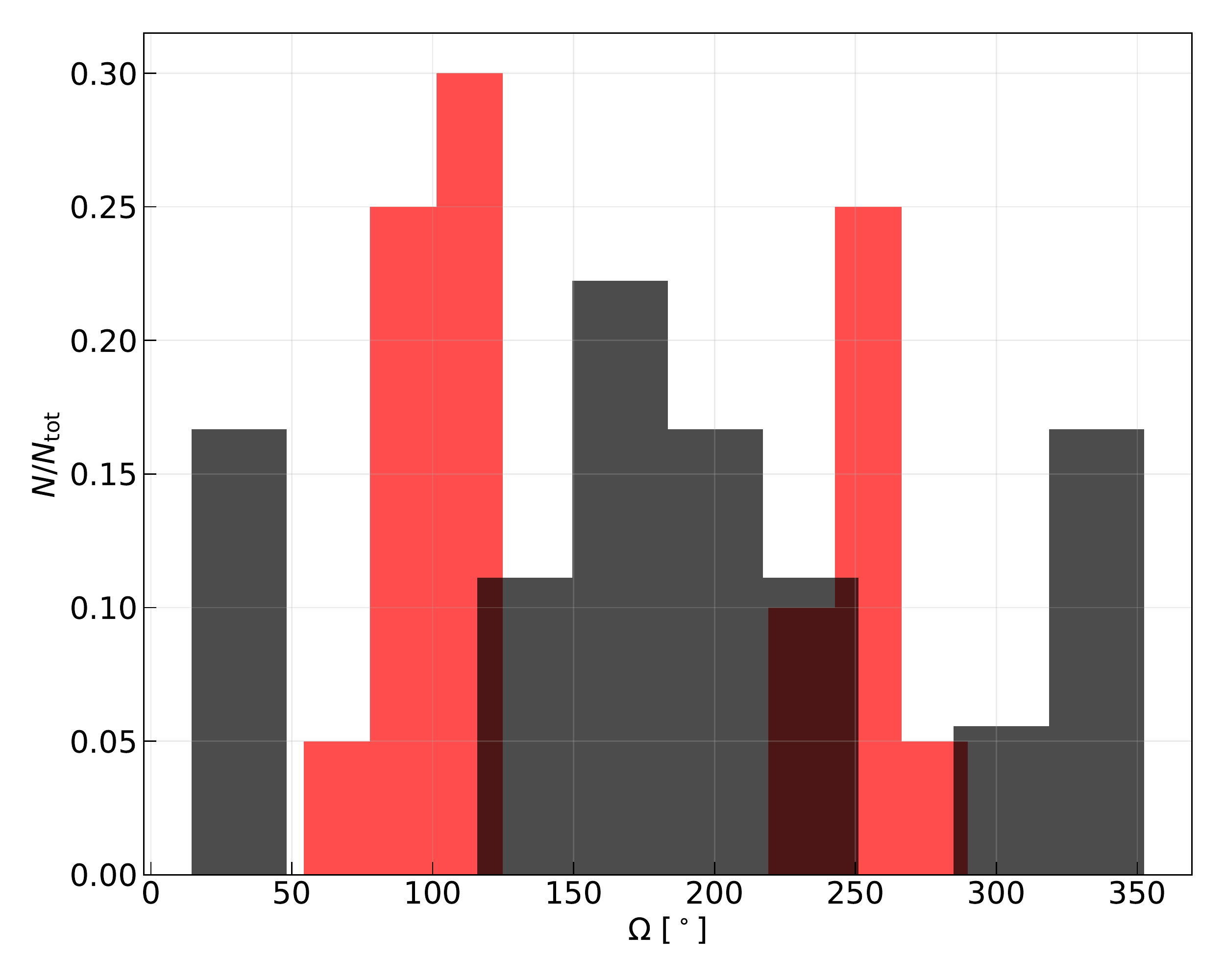} \\ 
		\par\end{center}
	\caption{Longitudes of the ascending nodes of the observed $S$-stars with known orbital elements. The  black and red histograms represent the black and red discs according to \citet{Ali2020}. }
	\label{fig:s-lans}
\end{figure}

Recent observations of the $S$-stars\footnote{Here we define $S$-stars as all the stars in the Galactic Centre with known full orbital solutions around the SMBH as reported by \citet{Ali2020} and \citet{Peissker2020}.} in the Galactic centre revealed that the kinematic structure of the stars with known orbital parameters appears to resemble two orthogonal discs \citep{Ali2020, Peissker2020} labelled as ``red'' and ``black'' discs. The discs can be identified from the distribution of the position angles of the semimajor axes projected on the sky which in turn is reflected in the distribution of the longitudes of ascending nodes (LaNs) of the orbits. Fig. \ref{fig:s-lans} shows the distribution of LaNs of the black and red discs in the form of two normalised histograms separately for each of the discs as classified by \citet{Ali2020}. The peaks around 0, 180 and 360$^\circ$ correspond to one plane of the black disc while two peaks around 100 and 270$^\circ$  show that the red disc is almost orthogonal to the black one. 

We compare the observed properties of the $S$-stars with the orbital parameters in three of our 1X simulations: \textit{stardisc} $\gamma=3.3$, \textit{stardisc} $\gamma=2.4$ and the \textit{thermal} model (see Table~\ref{tab:runs}). We examine the simulation snapshots at 5 Myr. The \textit{stardisc} initial conditions represent the case when the stars formed from the fragmenting gaseous accretion disc and the stars residing inside 0.05 pc migrated from the outer regions due to gas-driven planetary-type migration \citep{Levin2007}. This leads to nearly circular orbits matching the \textit{stardisc} initial conditions. Alternatively, massive stars could form by accreting matter from AGN discs \citep{Levin2007,Davies2020, Cantiello2021}. Another way to form the disc of stars is by disruption of a molecular cloud resulting in high orbital eccentricities (see e.g. \citealt{Generozov2021}). This formation scenario is closer to our \textit{thermal} model. 

\begin{figure}
	\begin{center}
		\includegraphics[width=\columnwidth]{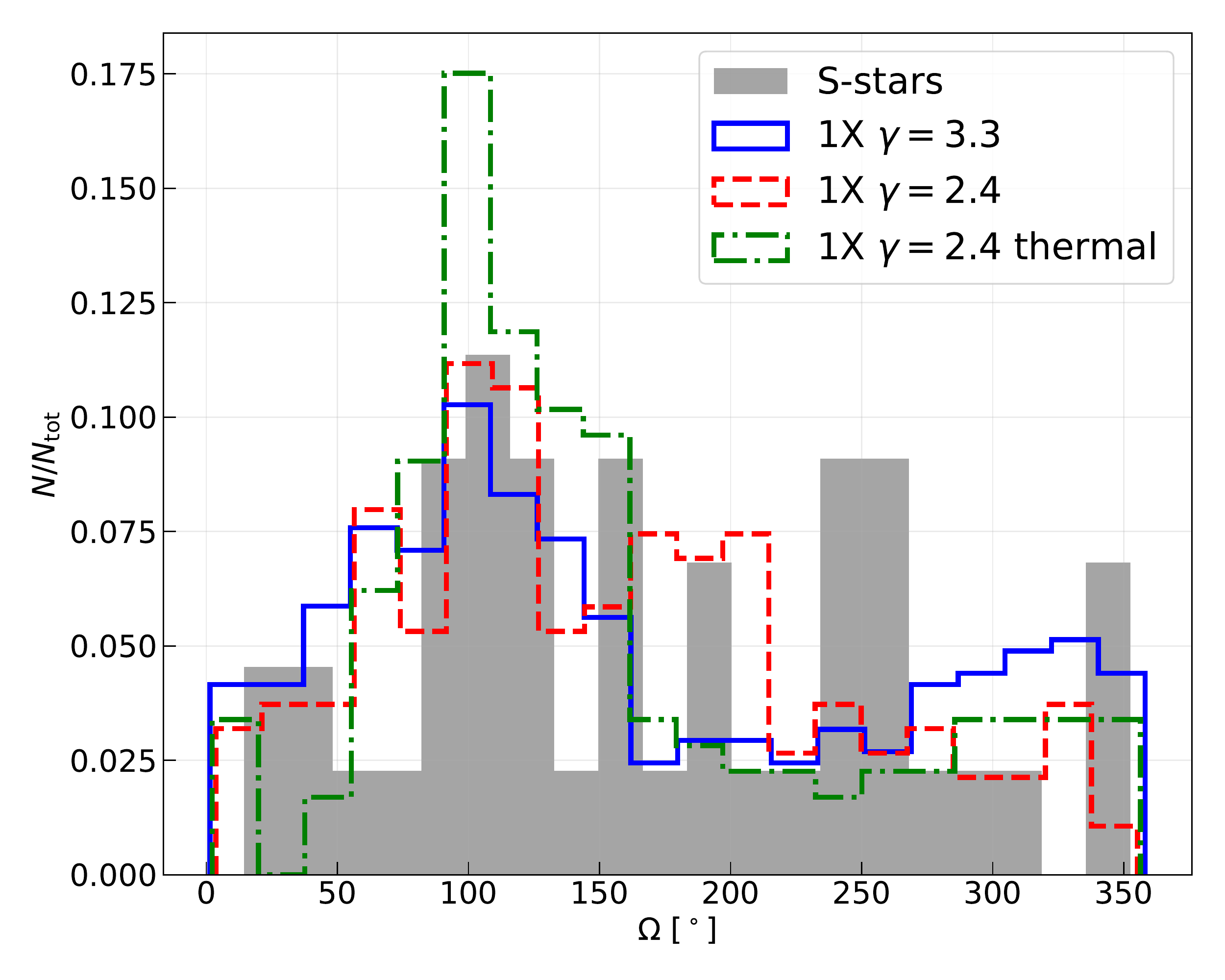} \\ 
		\par\end{center}
	\caption{Longitudes of the ascending nodes for the inner stars (a < 0.05 pc) in simulations at 5 Myr (blue, red and green lines) compared to the $S$-stars (shaded). The histogram for the $S$-stars shows both red and black discs. The reference direction for the longitudes of the ascending nodes in the simulations is chosen to match the peak in $S$-stars.}
	\label{fig:s-lans-comp}
\end{figure}

\begin{figure}
	\begin{center}
		\includegraphics[width=\columnwidth]{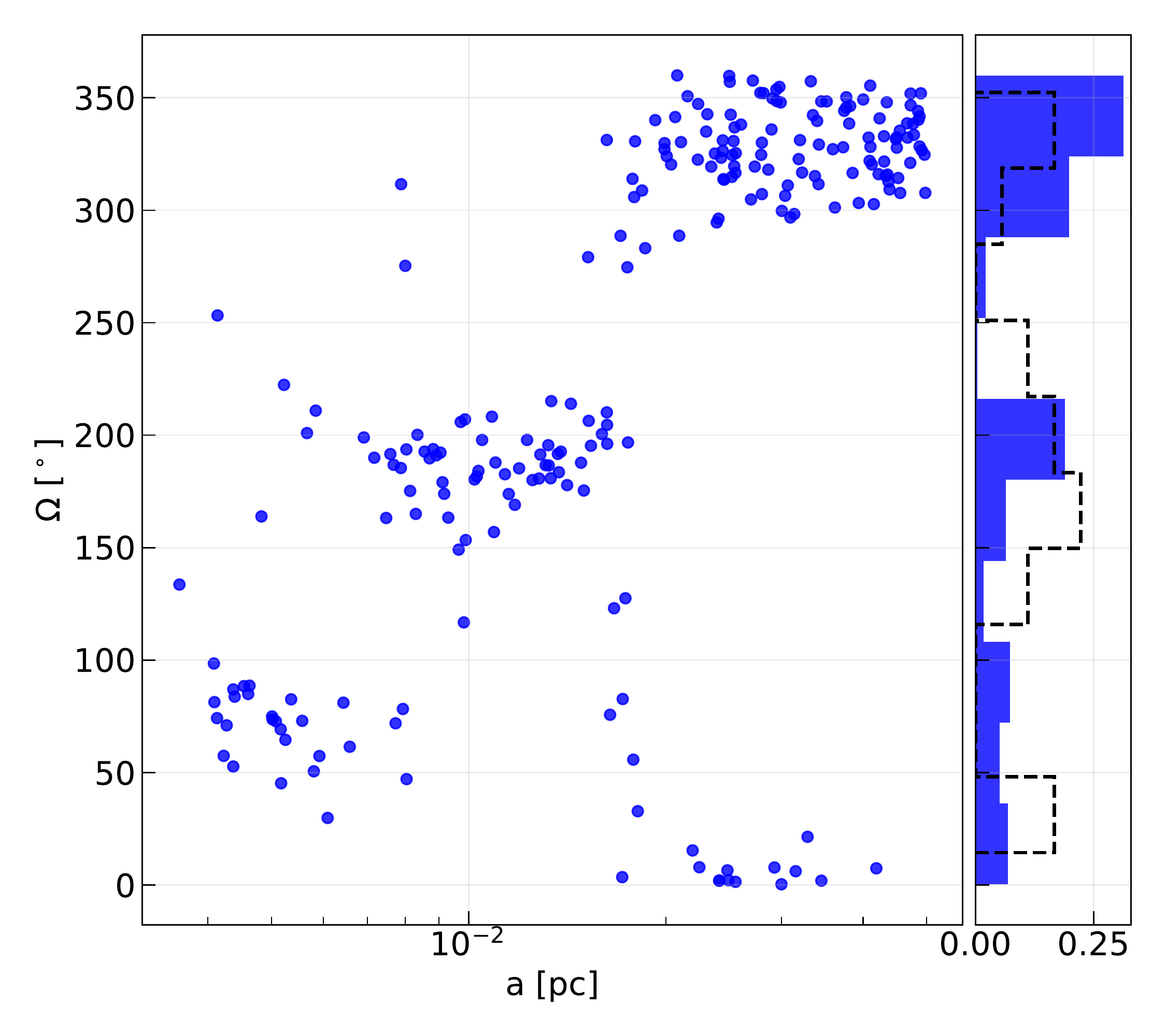} \\ 
		\par\end{center}
	\caption{Correlation of the longitudes of the ascending nodes for the inner stars (a < 0.05 pc) with semi-major axes for the model \textit{stardisc} $\gamma=2.4$ at 1.1 Myr. Right panel shows the corresponding normalised histogram where the dashed line represents the distribution of stars from the "black" disc from \citet{Ali2020}. }
	\label{fig:a-lans}
\end{figure}

\begin{figure}
	\begin{center}
		\includegraphics[width=\columnwidth]{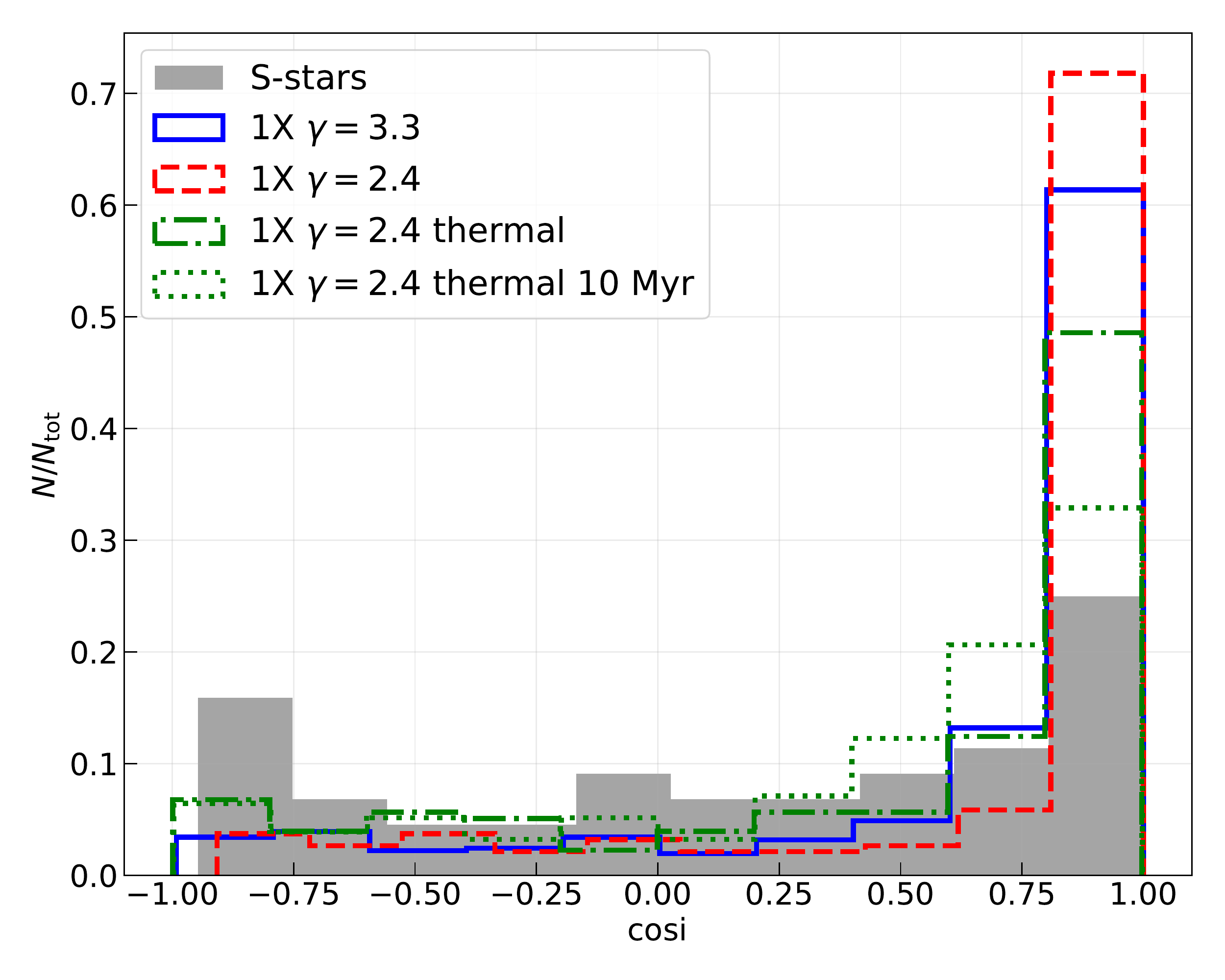} \\ 
		\par\end{center}
	\caption{Distribution of cosines of inclination angles of stars with respect to the principal eigenvectors. Shaded histogram corresponds to the $S$-stars and the coloured histograms show stars in the inner region of the stellar disc (a < 0.05 pc) from the simulations at 5 Myr (except for the green dotted line which corresponds to 10 Myr).}
	\label{fig:s-cosi}
\end{figure}

\begin{figure}
	\begin{center}
		\includegraphics[width=\columnwidth]{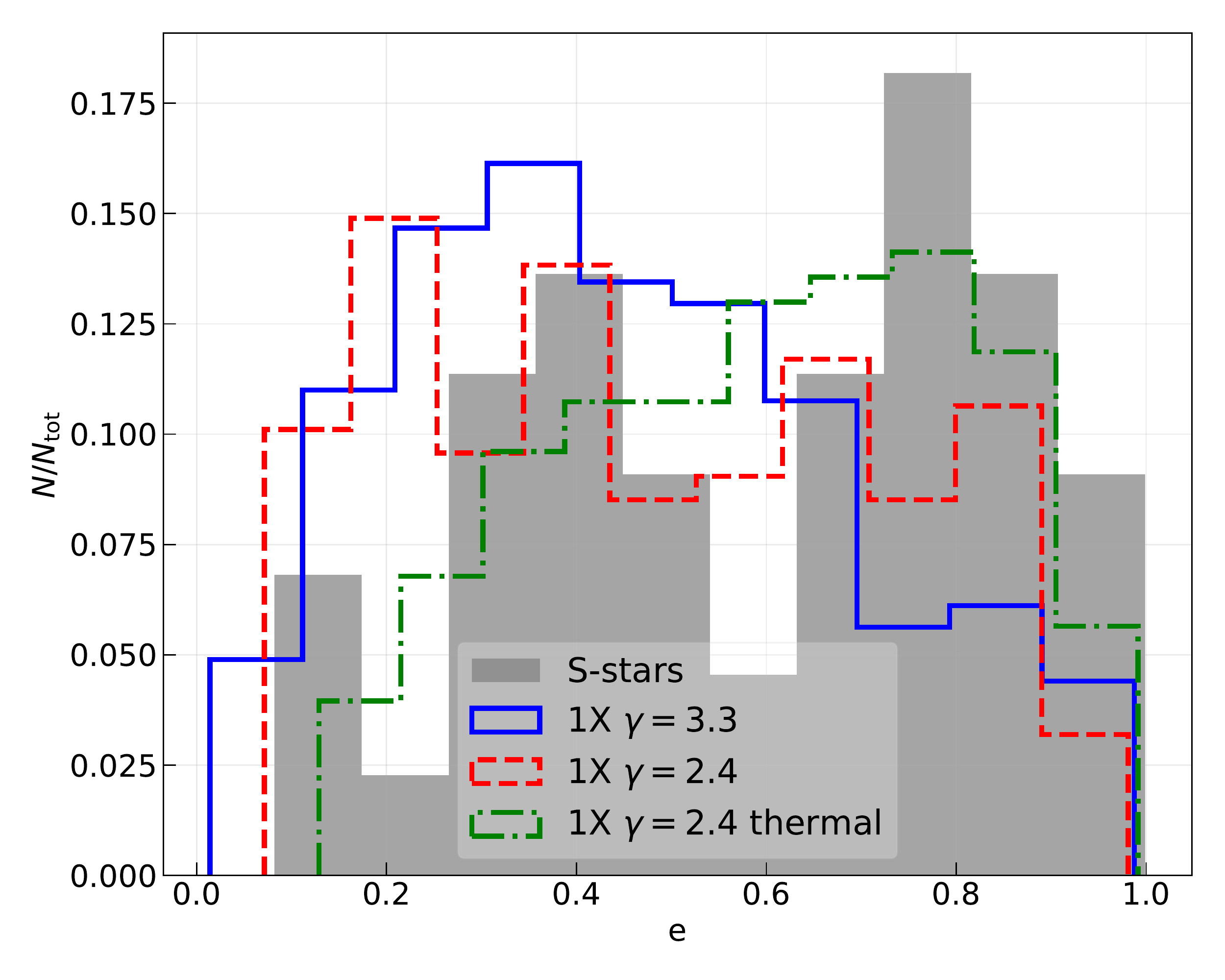} \\ 
		\par\end{center}
	\caption{Eccentricity distributions for the inner stars (a < 0.05 pc) in simulations at 5 Myr (blue, red and green lines) compared to the $S$-stars (shaded).}
	\label{fig:s-ecc}
\end{figure}

To compare the observed distribution of inclination angles of the $S$-stars we convert the data provided by \citet{Ali2020} and \citet{Peissker2020} to the coordinates with respect to the principal eigenvector of the system (Eq.~\ref{eq:Eigenvectors}). This way the inclination angles are independent of the choice of the reference plane of the coordinate system. To define the longitude of ascending nodes in our simulations, we orient the $x-y$ axes such that the peak of the distribution matches that of the S-stars. We select stars from the inner region of the stellar disc ($a<0.05\,$pc) and compare their properties to the observational data of the $S$-stars. Due to the observational limits, only the stars with masses $m \geq 3M_\odot$ can be detected, however we did not use the mass criterion to select the $S$-stars from our simulations. This is because we previously showed that stellar discs embedded in a spherically symmetric and isotropic stellar component have no vertical or eccentric mass segregation. If in reality the distribution of low mass $S$-stars will be different (when they are detected) from high mass $S$-stars, this would point to a larger amount of initial anisotropy of the background (old) stellar population surrounding the $S$-stars than assumed in our models.

We start by comparing LaNs (Fig.~\ref{fig:s-lans-comp}). The shaded histogram in Fig.~\ref{fig:s-lans-comp} shows the observed $S$-stars without dividing them into two discs. As we can see, all three of our models feature a peak around $100^\circ$ matching with the $S$-stars by construction. This anisotropy is caused by the fluctuating torques from the spherical component. However, we cannot clearly detect the second peak corresponding to another disc (black disc) nor the opposite peak corresponding to a counter-rotating component of the disc (red disc) in the same plane ($\Omega \sim 250^\circ$). 

We note that two distinct peaks in LaNs form in the \textit{stardisc} $\gamma=2.4$ model at 1.1 Myr, but this feature is transient and dissolves in less than 0.5 Myr. Fig.~\ref{fig:a-lans} shows a scatter plot of LaNs versus semi-major axes for this model indicating that each peak in the distribution correspond to different semi-major axes. Comparison with the data from \citet{Ali2020} yields similar properties with the ``black" disc (shown as a dashed line in the histogram in Fig.~\ref{fig:a-lans}). 

Fig.~\ref{fig:s-cosi} shows the distribution of the cosines of the orbital inclinations with respect to the principal eigenvector in the observations and in our simulations. The observed red disc and black discs correspond to the peaks at $\cos i=\pm 1$ and at 0, respectively. In contrast, the simulations have a more prominent peak at $\cos i=1$ and do not show a peak at $\cos{i} \simeq -1$ indicating a lack of retrograde stars in the same plane. Furthermore, the simulations do not display a peak at $\cos i = 0$. Our models also show significantly higher relative number of stars in the mid-plane of the disc ($\cos{i}=1$), indicating less diffusion took place from the initial condition in the simulations than observed. This suggests that the orbits of the observed $S$-stars are at a later stage of angular momentum relaxation. The dotted line in Fig.~\ref{fig:s-cosi} shows the distribution of cosines of orbital inclinations for the \textit{thermal} model at 10 Myr. Because the \textit{thermal} model is the most efficient in terms of VRR (see also Appendix~\ref{app:RR}), this implies that even 10 Myr is not enough to fully randomise the orbital inclinations. 

Massive perturbers such as a cusp of stellar black holes or an intermediate-mass black hole (IMBH) may boost both two-body and resonant relaxation \citep{Perets2007,Kocsis2011,Kocsis2015}. Let us estimate the mass of an IMBH required to speed up VRR by a certain factor $\kappa$. Following Eq.~\eqref{eq:t-vrr} and applying the definition of the effective mass, gives
\begin{equation}
    m_\mathrm{IMBH} = \sqrt{\left(\kappa^2-1\right)\sum_{i=1}^N m_\mathrm{i}^2} = \left(\kappa^2-1\right)^{1/2} N^{1/2} \langle m^2\rangle^{1/2},
\end{equation}
where $m_\mathrm{i}$ is the mass of $i^\mathrm{th}$ star and $N$ is the total number of stars (in our case within 0.05 pc), and $N\simeq4500$ and $\langle m^2\rangle^{1/2}=5.33\msun$ in our models in this region. For example, to speed up VRR by a factor of $\kappa=2$ one needs an IMBH of $m_\mathrm{IMBH}\simeq620\msun$. In Appendix~\ref{app:RR} we show that VRR for a stellar disc embedded in a spherical component is quenched by a factor $\beta_\mathrm{T,s}/\beta_\mathrm{T,d}$. For the \textit{stardisc} $\gamma=2.4$ model, to speed up VRR so that an IMBH balances the quenching from the disc one needs $\kappa=\beta^2_\mathrm{T,s}/\beta^2_\mathrm{T,d}\simeq2.23$ , i.e. an IMBH of $m_\mathrm{IMBH}\simeq710\msun$.
Under certain conditions an IMBH may also produce counter-rotating stars in the same plane and give rise to a second stellar disc (Panamarev, Zou, Kocsis, in preparation).

Finally, Fig.~\ref{fig:s-ecc} shows the distribution of eccentricities, indicating that the observed sample of S-stars exhibits two distinct peaks near 0.4 and 0.8 \citep{Ali2020}. In contrast, neither of our simulations show two peaks, but interestingly the \textit{stardisc} models match the peak at $e=0.4$ while the \textit{thermal} model matches the peak at $e=0.8$. However, note that the observed sample of $S$-stars from \citet{Ali2020} contains only a small sample of $\sim40$ stars where the significance of the two peaks are greatly decreased by Poisson fluctuations. 

Thus, if the $S$-stars formed in a disc, the simulations suggest that the distribution of their orbital angular momentum vectors should have retained a stronger peak up to at least 10 Myr since their formation, and to match the observed distribution the root-sum-squared mass in the same region should be $(\sum_i m^2)^{1/2} = 820\msun$ which is possible with an initial stellar disc of $N_{\rm d}=10^3$ stellar objects and remnants and an IMBH of mass $500-1000\msun$, or with a massive cusp of stellar black holes.

\section{Summary and Discussion} 
\label{sec:SUM}

We performed a set of direct $N$-body simulations of nuclear stellar discs with a massive black hole at the centre. We examined cases with and without a spherical star cluster in the same region. We presented the first one-to-one direct $N$-body simulations of the inner 0.5 pc of the Milky Way nuclear star cluster featuring a realistic total stellar mass and a top-heavy mass function. Furthermore, we ran simulations which represent the conditions at the centres of ultracompact dwarf galaxies. Our main findings are as follows.
\begin{itemize}
    \item The relaxation processes in isolated stellar discs lead to vertical and eccentric mass segregation meaning that massive stars settle to lower orbital inclinations and more circular orbits than the light stars. This is caused by both resonant and two-body relaxation. On the other hand, the interaction with an isotropic spherical distribution of stars quenches mass segregation in inclinations and eccentricities. 
    \item The interaction of a stellar disc with a spherical component leads to the thickening of the stellar disc. The rate of this process depends strongly on the semimajor axis. The stars in the inner region relax faster in terms of inclination angles leading to a anticorrelation between orbital inclinations and the distance from the SMBH. Our simulations showed that for conditions in the Milky Way, the orbital inclinations change predominantly due to VRR, despite the fact that VRR is quenched by nodal precession due to the torques from within the stellar disc.
    \item The nuclei of dwarf galaxies hosting stellar discs and massive black holes of order $10^5 \msun$ are dominated by two-body relaxation.  
    These systems approach full mixing on the two-body relaxation timescale, where an initially thin disc becomes spherical if embedded in a much more massive spherical cusp. The spherical component does not develop a significant flattening if the disc mass is less than 15\% of the spherical cluster, but very massive discs (comparable with the mass of the sphere and more massive) cause flattening of the initially spherical distribution and drive anisotropic mass segregation.
    \item The dynamics of the $S$-stars at the Galactic centre from their formation up to 5 Myrs is dominated by VRR. This results in an anticorrelation of orbital inclinations with distance from the SMBH meaning that the thickness of the disc increases with decreasing radius which is confirmed in recent observations \citep{vonFellenberg2022}. 
    The stochastic deviations from an isotropic distribution in the spherical component of old stars gives rise to a non-zero net torque which leads to 
    an overdensity of angular momentum vectors in a given direction, hence a
    peak in distribution at a particular value. However, this does not explain the distribution of longitudes of the ascending nodes presented by \citet{Ali2020} which they interpret as two orthogonal counter-rotating discs.
    \item 
    Our simulations led to less diffusion of angular momentum vector directions from a thin stellar disc in 10 Myr than currently observed for the $S$-stars. This suggests that if the $S$-stars initially formed in a stellar disc, the root-sum-squared mass of stellar objects and remnants in this region should be of order $820\msun$ within 0.05pc to reproduce the observed scatter at present in angular momentum vector directions, suggesting that the $S$-stars co-exist with a cusp of stellar black holes or with an IMBH of mass $m_\mathrm{IMBH} = 500-1000\msun$ (see \citealt{Gravity2020} and references therein for limits on an IMBH in the Galactic centre).
\end{itemize}

Our simulations of the inner part of the Milky Way nuclear star cluster featured a realistic number of stars within 0.5 pc, but one of the assumptions for the spherical stellar component was a nearly exactly isotropic distribution of angular momentum vectors (deviations at the level of $10^{-4}$) which is expected to be responsible for the absence of vertical mass segregation in our models. Thus, one of the next steps to explore the evolution of stellar nuclear discs is to study the interaction with stellar systems with anisotropy and/or rotation. This is reasonable as observations show that the Milky Way nuclear star cluster has net rotation and flattening \citep{Feldmeier2014}. Moreover, recent observations suggest that 7\% of the stars in the inner parsec exhibit faster rotation \citep{Do2020, Arca-Sedda2020}. Theoretical studies of the VRR indicate that initial anisotropy in the distribution of stellar angular momenta strongly affect the final equilibrium distribution of multi-mass stellar systems \citep{Szolgyen2018, Mathe2022, Magnan2021}. Furthermore, $N$-body simulations of rotating globular clusters show that vertical mass segregation may also occur in globular clusters \citep{Szolgyen2019,Tiongco2021, Tiongco2022}.

The explored initial conditions included the results of previous stardisc simulations of active galactic nuclei \citep{Panamarev2018} leading to relatively old stellar population within the disc, but this is not the case in the Galactic centre \citep{LevinBeloborodov2003}. One way to form young stars matching the initial conditions explored in this paper is to form the stars from the gaseous accretion disc. This type of formation scenario was studied by \citet{Levin2007} predicting innermost stars on circular orbits. One way to improve our models and to account for young stars would be to perform simulations with stellar evolution assigning two different populations for the disc and the sphere. As the stellar evolutionary mass loss is high for the most massive stars, this may affect the resulting kinematic signatures of massive stars.

We did not take into account the effect of the outer galaxy in the simulations. This is justified because we modelled the innermost part of the galactic nucleus where the potential is highly dominated by the SMBH, while the contribution from the galactic components like bulge, disc or halo becomes important at larger scales, outside the influence radius of the SMBH.

The configuration of the \textsc{$\varphi$-grape} code used in this study was designed to avoid formation of binary stars in the explored stellar systems. But it was shown that binary stars may significantly alter the observed orbital elements of the stars in the young stellar disc at the Galactic centre \citep{Naoz2018}. Moreover, one of the formation scenarios of the $S$-stars is the Hills mechanism which involves tidal disruptions of binaries by the SMBH \citep{Hills1975,Perets2007, Fragione_Sari2018,Generozov2021}. $S$-stars formed as a result of the Hills mechanism are expected to feature initially high eccentricities contrary to the in-situ formation studied in this paper. Therefore, a next step to improve our models is to incorporate formation and evolution of binaries starting with the stellar disc with a fraction of stars in binary systems. Moreover, this will allow us to study the effect of binaries on the efficiency of resonant relaxation processes in galactic nuclei hosting stellar discs. Simulations including binaries with and without stellar evolution may be done using \textsc{nbody6++gpu} code \citep{Wang2015} with the most recent updates of the stellar evolution \citep{Kamlah2022a,Kamlah2022b}. 

Another way to improve our models is to combine direct $N$-body modelling with self-consistent field models \citep{Meiron2014} to account for the dynamical effects of the embedding galaxy on the nuclear stellar disc. An example of this approach is the direct integration of all the disc particles and the innermost particles in the sphere (e.g. within 0.1 pc) and hybrid integration of the outer stars (r > 0.5 pc). This would speed up the simulations and allow to reach larger masses and number of particles (up to $10^6$) within the inner 0.5 pc in the Milky Way and potentially to model nuclei of more massive galaxies hosting SMBHs, nuclear star clusters and stellar discs.

\section{Acknowledgements}
We thank Peter Berczik for assistance with the \textsc{$\varphi$-grape} code and for useful comments and suggestions.
This work received funding from the European Research Council (ERC) under the European Union’s Horizon 2020 Programme for Research
and Innovation ERC-2014-STG under grant agreement No. 638435 (GalNUC). This work was supported by the Science and Technology Facilities Council Grant Number ST/W000903/1. We acknowledge the support of  the  Science Committee of the Ministry of Education and Science of the Republic of Kazakhstan (Grants No. AP08856184 and AP08856149).

\section*{Data Availability}
The data underlying this article will be shared on reasonable request to the corresponding author.

\bibliography{refs}

\appendix

\section{Measuring the efficiency of relaxation processes}
\label{app:RR}

\begin{figure*}
	\begin{center}
		\includegraphics[width=\linewidth]{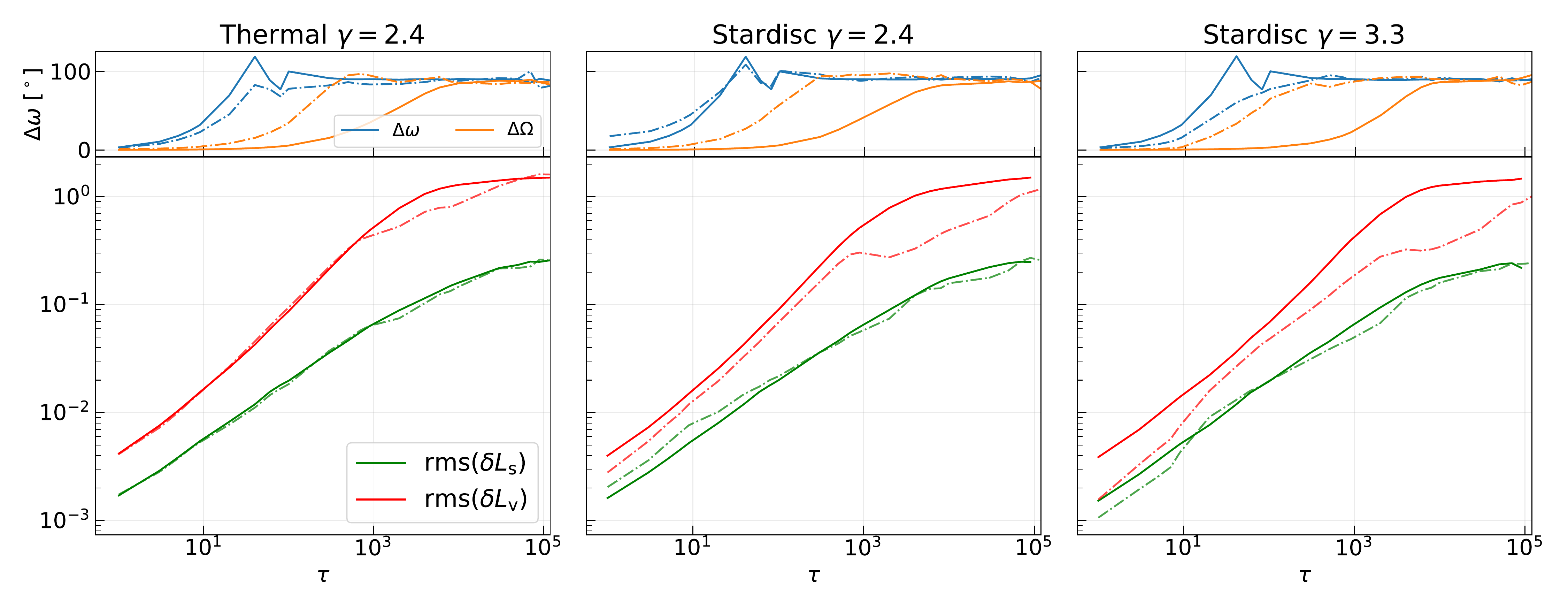} \\ 
		\par\end{center}
	\caption{Similar to Fig.~\ref{fig:rr-od} and Fig.~\ref{fig:rr} but only for 1X models with a spherical component. Left panel shows the model with \textit{thermal} disc with $\gamma=2.4$, mid and right panels show the \textit{stardisc} models with $\gamma=2.4$ and $\gamma=3.3$ respectively. The dash-dotted lines on all panels show the curves corresponding only to the stars that initially belong to the disc; solid lines show values for the stars that initially belong to a spherical component. Upper panels show change in argument of periapsis (which demonstrates the apsidal precession rate) and longitude of the ascending node (demonstrates the nodal precession rate). } 
	\label{fig:rr_1x}
\end{figure*}

To measure the efficiency of relaxation processes we follow the steps described in \citet{Rauch1996, Eilon2009, Meiron2019}. We compute the relative change in Keplerian energy, angular momentum vector magnitude, angular momentum vector direction and $z$-component of the angular momentum vector defined as:

\begin{align}
    \delta E = \frac{E - E_0}{E_0},\;
    \delta L_s = \frac{|\boldsymbol{L}| - |\boldsymbol{L}_0|}{L_{\rm c}},\;
    \delta L_v = \frac{|\boldsymbol{L} - \boldsymbol{L}_0|}{L_{\rm c}},\;
\label{eq:dLz}
    \delta L_z = \frac{|\boldsymbol{L}_z - \boldsymbol{L}_{z,0}|}{|\boldsymbol{L}_{z,0}|}
\end{align}
with respect to the normalised time defined as:
\begin{equation}
\label{eq:tau}
    \tau = \frac{t - t_0}{t_\mathrm{orb}}.
\end{equation}
Where $t_0$ is the initial moment in time which was chosen to correspond to $t=0$, $E_0$, $L_0$ and $L_{z,0}$ correspond to the time $t = t_0$. 
The defined above quantities are computed for each particle corresponding to a relevant bin in the normalised time $\tau$. After that, we compute rms for all particles in each bin. The rms of energies and angular momenta are plotted in Fig.~\ref{fig:rr-od} and Fig.~\ref{fig:rr}.

Together with changes in energies and angular momenta, we track changes in the arguments of periapsides and the longitudes of the ascending nodes.
\begin{equation}
    \delta\omega = \arccos{(\omega -\omega_0)}\,,\quad
    \delta\Omega = \arccos{(\Omega -\Omega_0)}\,.
\end{equation}
The top panes in Fig.~\ref{fig:rr-od} and Fig.~\ref{fig:rr} show the mean change in these quantities with respect to $\tau$.

To measure the rate of relaxation, we assume the following relations \citep{Rauch1996, Meiron2019}:
\begin{align}
    \mathrm{rms}(\delta E) &= \alpha\frac{m_2}{M_\mathrm{bh}}\sqrt{N}\sqrt{\tau}\,,\\
    \mathrm{rms}(\delta L_s) &= \eta_s\frac{m_2}{M_\mathrm{bh}}\sqrt{N}\sqrt{\tau}\,,\\
\label{eq:rms_lv}
    \mathrm{rms}(\delta L_v) &= \frac{m_2}{M_\mathrm{bh}}\sqrt{N}(\eta_\nu\sqrt{\tau} + \beta_\nu\tau)\,.
\end{align}

We focus on the coherent part of VRR, where the efficiency is linear with $\tau$ and is given by (following Eq.~\ref{eq:rms_lv}):
\begin{equation}\label{eq:beta_nu}
    \beta_\nu =  \frac{d\,\mathrm{rms}(\delta L_\mathrm{v})}{d\tau}\frac{M_\mathrm{bh}}{m_2\sqrt{N}}  
\end{equation}
The definition of $\beta_\nu$ is somewhat different in different studies \citep{Rauch1996, Gurkan2007, Eilon2009, Kocsis2015}, we use the definition of \citet{Kocsis2015} where $\beta_\nu$ is replaced by:
\begin{equation}\label{eq:beta_t}
    \beta_\mathrm{T} = \frac{ \beta_\nu m_2 }{ \mathrm{rms}(m) \sqrt{3-\gamma} },  
\end{equation}
where $\gamma$ is the power-law density slope of the system, $\mathrm{rms}(m)=4.95\msun$ is the root-mean-square of stellar masses and $m_2=12.17\msun$ is the effective mass.

In Fig.~\ref{fig:rr_1x}, the coherent phase of VRR is clearly seen in the range of $100 < \tau < 500$. We perform a linear fit in this range. 

\begin{table}
\caption{Measured values of $\beta_\mathrm{T}$ and $\beta_\mathrm{EKA}$ for simulations with a disk and spherical component}
\label{tab:beta}
\begin{center}
\begin{tabular}{lcccc}
\hline\hline
 & sphere & disc  & disc  &  disc \\
 &  & stardisc  &  stardisc  & thermal  \\
$\gamma$ & 1.75 & 3.3  &  2.4  & 2.4   \\
\hline
$\beta_\mathrm{T}$  & 1.44 & 0.44 & 0.96 & 1.33 \\
$\beta^\mathrm{EKA}$  & 0.99 & 0.30 & 0.66 & 0.92 \\
\hline\hline
\end{tabular}
\end{center}
\textbf{Notes.} List of the values for $\beta_\mathrm{T}$ for the spherical and disc components in different models for the discs according to the definition from \citet{Kocsis2015} in comparison with the definition used in \citet{Eilon2009}. First column shows the value obtained for the stars in a spherical component, the remaining columns indicate the values for the disc stars corresponding to different disc models with different $\gamma$ radial density profile exponents. 
\end{table}

To measure the effect of the stellar disc on the efficiency of VRR, we measure $\beta_\mathrm{T}$ (Eq.~\ref{eq:beta_t}) separately for the stars that are initially arranged in the disc ($\beta_\mathrm{T, d}$) and for the stars that belong initially to the spherical component ($\beta_\mathrm{T, s}$). However for a crude estimate, in both cases we use the rms mass and $m_2$ and $\gamma$ factor of the spherical component in Eq.~\eqref{eq:beta_t} even when calculating $\beta_\mathrm{T, d}$ since the spherical component is expected to dominate the evolution of disc stars. We perform the measurement of $\beta_\mathrm{T, d}$ and $\beta_\mathrm{T, s}$ for three 1X models that we used to compare with the $S$-stars in Sec.~\ref{sec:s-stars}: \textit{thermal}, \textit{stardisc} $\gamma=2.4$ and \textit{stardisc} $\gamma=3.3$. As a result, we find that for the \textit{stardisc} models VRR is more efficient for stars in the spherical component, while for the stars that initially reside in the disc, VRR is less efficient. As we see from Fig.\ref{fig:rr_1x}, the coherent accumulation of torques of VRR is limited by the nodal precession time (orange lines in top panels) after which we see a random walk growth. This is clearly seen for the disc stars (dashed-dotted lines in Fig.~\ref{fig:rr_1x}). Note that the least efficient VRR regime is in the \textit{stardisc} $\gamma=3.3$ model, where VRR is quenched by a factor of $\beta_\mathrm{T, s}/\beta_\mathrm{T, d} \simeq 3.27$. This is explained by the fact that due to the steep density profile, the inner part of the whole system is largely dominated by the stellar disc leading to fast nodal precession rate. The model \textit{stardisc} $\gamma=2.4$ slows down the vector angular momentum relaxation rate by a factor of 1.5, but VRR in the \textit{thermal} model is quenched only by $\simeq$10\%. We summarise the measured values for $\beta_\mathrm{T, d}$ and $\beta_\mathrm{T, s}$ in Table~\ref{tab:beta} and compare them with $\beta^\mathrm{EKA}$ -- the definition of $\beta_\nu$ used in \citet{Eilon2009} which is related to $\beta_\mathrm{T}$ as $\beta_\mathrm{T} = 0.69\beta^\mathrm{EKA}$ \citep{Kocsis2015}.

\end{document}